\documentclass[aps,prd,showpacs,amssymb,amsmath,amsfonts,
twocolumn,floatfix,preprintnumbers]{revtex4}

\usepackage{graphicx}
\usepackage[raggedright,hang]{subfigure}
\usepackage{epstopdf}
\usepackage{rotating}
\usepackage{color}
\usepackage{hyperref}
\usepackage{amssymb}
\usepackage{amsmath}
\usepackage{fancyhdr}

\newcommand{\rme}{\mathrm{e}}

\newcommand{\rmd}{\mathrm{d}}
\renewcommand{\vec}[1]{\boldsymbol{\mathrm{#1}}}
\newcommand{\Or}{\mathord{\mathcal{O}}} 
\newcommand{\eref}[1]{(\ref{#1})}

\newcommand{\Eref}[1]{Equation~(\ref{#1})}

\newcommand{\Sref}[1]{Section~\ref{#1}}
\newcommand{\Fref}[1]{Figure~\ref{#1}}
\newcommand{\Tref}[1]{Table~\ref{#1}}
\newcommand{\Hz}{\mathrm{Hz}}
\def\SSB{\textrm{\mbox{\tiny{SSB}}}}
\def\erf{\mathrm{erf}}
\def\sinc{\mathrm{sinc}}
\def\sig{\textrm{\mbox{\tiny{S}}}}
\newcommand{\fkdot}[1]{f^{(#1)}}
\newcommand{\xil}[1]{{\vec{\xi}}^{\,(#1)}}
\newcommand{\xia}[1]{{{\xi}}^{\,(#1)}}
\def\Doppler{\lambda}
\def\vDoppler{\Doppler}

\newcommand{\F}{\mathcal{F}}
\def\Xsym{\star}
\newcommand{\Xstat}{\mathcal{F}^\Xsym}
\newcommand{\Xamp}{\mathcal{X}}
\newcommand{\Xest}{\mathrm{X}}
\newcommand{\Hyps}{\mathcal{H}}

\newcommand{\Pspace}{\mathcal{P}}
\newcommand{\Uspace}{\mathcal{U}}
\newcommand{\vecu}{\mathrm{u}}

\DeclareGraphicsExtensions{.eps}

 \def\dcc{LIGO-P080051-02-Z}
 \def\aei{AEI-2008-043}

\begin{document}

\pagestyle{fancy}

\preprint{\dcc}
\preprint{\aei}

\rhead[]{}
\lhead[]{}

\title{Parameter-space correlations of the optimal statistic\\
for continuous gravitational-wave detection}

\author{Holger~J.~Pletsch}
\email{Holger.Pletsch@aei.mpg.de}
\affiliation{Max-Planck-Institut f\"ur Gravitationsphysik (Albert-Einstein-Institut)
    and Leibniz Universit\"at Hannover, Callinstr. 38, 30167 Hannover, Germany}

\begin{abstract}
The phase parameters of matched-filtering searches for
continuous gravitational-wave signals are sky position,
frequency and frequency time-derivatives. The space of these
parameters features strong global correlations in the optimal detection
statistic.  For observation times smaller than one year, the orbital motion of
the Earth leads to a family of global-correlation equations which
describes the ``global maximum structure" of the detection 
statistic. The solution to each of these equations 
is a different hypersurface in parameter space.
The expected detection statistic is maximal at  the intersection of 
these hypersurfaces. The global maximum structure of the 
detection statistic from stationary instrumental-noise 
artifacts is also described by the global-correlation equations. 
This permits the construction of a  veto method which excludes 
false candidate events.
\end{abstract}


\maketitle

\section{Introduction\label{sec:Introduction}}

The emission of continuous gravitational waves (CW) is expected, for
instance, from spinning neutron stars with non-axisymmetric
deformations.  If the system is isolated, it is losing angular
momentum through radiation and is slowing down. Therefore the
gravitational-wave frequency would be slowly decreasing for this
long-lasting type of signal.  Such CW sources are among the primary
targets of Earth-based, laser-interferometric and resonant-bar
detectors.

The terrestrial location of the detectors generates a Doppler modulation 
of the signal caused by the detector's motion relative to the solar system 
barycenter~(SSB). The observed phase therefore depends on ``phase parameters",
which describe the intrinsic frequency evolution and the source's sky location.  
In addition, there is a time-varying amplitude modulation due to the 
antenna patterns changing with the Earth's spinning motion.  
The latter variations depend on the
``amplitude parameters", which are the two polarization amplitudes,
and the polarization angle of the gravitational wave.

To extract CW signals buried in the detector noise, 
the optimal data analysis scheme is derived in~\cite{jks1} 
based on the principle of maximum likelihood detection leading to 
coherent matched filtering. It is shown, that the amplitude parameters together with
the initial-phase parameter can be eliminated by analytically
maximizing the detection statistic, such that the search space is
just the phase parameters: sky position, frequency and
frequency time-derivatives.  This detection
statistic is commonly referred to as the $\F$-statistic.  For a given
sequence of data, wide-band all-sky searches evaluate the
$\F$-statistic over a large number of  template-grid points
in parameter space. The parameters of the templates for
which a predetermined threshold is exceeded are registered 
as candidate events for potential gravitational-wave signals.

In the {\it local} parameter-space neighborhood of a given signal one
can define a metric~\cite{{owen:1996me},{bccs1:1998},{prix:2007mu}} from 
the fractional loss in expected $\F$-statistic. 
This fractional loss defines the dimensionless ``mismatch"~$\mu$.
Let~$\vDoppler$ define a vector of phase parameters for a template. 
If~$\vDoppler_\sig$ denotes the signal's phase parameters, then 
the metric~$g_{ij}(\vDoppler_\sig)$ is obtained by Taylor-expanding the mismatch 
at $\vDoppler_\sig$ with respect to the small parameter offsets 
$\Delta\vDoppler \equiv \vDoppler_\sig - \vDoppler$ to quadratic order:
$\mu= \sum_{ij} \ g_{ij}(\vDoppler_\sig) \ \Delta \vDoppler^i \Delta \vDoppler^j
+ \Or[(\Delta \vDoppler)^3]$. 

This work identifies the {\it global} parameter-space
regions where the detection statistic~$\F$ is expected to have large values close to 
maximal without restriction to the local neighborhood of the signal location.
In this paper, these regions are referred to as the ``{\it global large-value structure}".
To find the global large-value structure of the $\F$-statistic,
a simplified detection statistic~$\Xstat$ (approximating $\F$) is considered.
The locations in parameter space where~$\Xstat$ is expected to be maximal 
are referred to as the ``{\it global maximum structure}".
For increasing parameter offsets from the given signal's parameters this 
global maximum structure of~$\Xstat$
(and therefore also the global large-value structure of $\F$) is found to become
significantly different from local approximation obtained from the metric.

A previous study~\cite{prixitoh:2005} examined monochromatic signals in the restricted
phase-parameter space of sky location and frequency. 
It was shown that in some different approximate detection statistic, 
such signals can generate ``circles in the sky" while searching a range of template frequencies.  
The collection of these circles forms a two-dimensional hypersurface in the three-dimensional
parameter space, described by a single equation.
Here, even in this restricted parameter space, 
this hypersurface is only an approximation of the description of the 
global maximum structure of the $\Xstat$-statistic to first order in observation time $T$.

\begin{figure}
	\includegraphics[width=0.8\columnwidth]
	{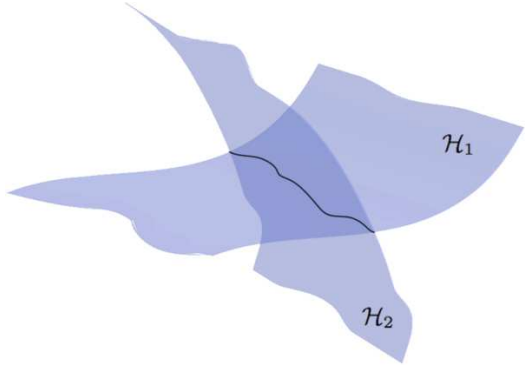}
	\caption{Schematic drawing of intersection of two representative 
	global-correlation hypersurfaces $\Hyps_1$ and $\Hyps_2$. 
	The region of intersection,
	shown by the solid black curve, describes the global maximum structure of the simplified 
	detection statistic~$\Xstat$, because the expectation value of~$\Xstat$ is maximal in these locations.
	For illustration purposes only two hypersurfaces are shown, whereas in general 
	one might need to consider more hypersurfaces. 
	\label{f:intersectdemo}}
\end{figure}

The present work shows that the global maximum structure of the detection
statistic~$\Xstat$ is described by a separate equation for each order of~$T$.
The solution to each of these equations is a different hypersurface in parameter space.
Therefore, it is this family of {\it global-correlation hypersurfaces},  which
describes the global maximum structure of~$\Xstat$:
The detection statistic~$\Xstat$ is expected to be maximal at the intersection of 
these hypersurfaces. This idea is illustrated schematically in \Fref{f:intersectdemo}.
The same results also apply when considering the generalization to
non-monochromatic signals allowing for an intrinsic frequency
evolution of the source.

This paper is organized as follows.
\Sref{sec:Matched filtering}  briefly reviews the
matched-filtering method for CW sources. 
\Sref{sec:simple-matched-filtering} describes an approximate
signal model which leads to the simplified detection statistic allowing the
analytical exploration of its global maximum structure.  
\Sref{sec:Global correlations} presents the global-correlation equations of the
phase-parameter space and illustrates the geometry of the
global-correlation hypersurfaces. In addition, the search-parameter regions 
are derived for which the approximations used are valid. 
In \Sref{sec:prediction-vs-Fstat} the predictions made
by the global-correlation hypersurfaces are compared to a fully
coherent $\F$-statistic search in data sets of software-simulated
sources with no detector noise as well as of hardware-injected CW signals
in the presence of real detector noise. 
In \Sref{sec:veto} the global large-value of the detection statistic which is caused by
stationary instrumental-noise artifacts mimicking astrophysical sources is
found to be also well described by the global-correlation
equations, and thus a veto method is constructed.  
\Sref{sec:Xspin} discusses the effects of the Earth's spinning motion
in the context of the present topic.
Finally a concluding section follows.

\section{Optimal filtering for continuous gravitational-wave signals
\label{sec:Matched filtering}}

In the absence of any signal, the detector output data time
series~$x(t)$ at detector time~$t$ only contains noise $n(t)$, which is assumed to be a
zero-mean, stationary and Gaussian random process. If a signal~$h(t)$
is present, the noise is assumed to be additive, $x(t) = n(t) + h(t)$.  
Denote by $t_{\SSB}$ the time measured at the solar
system barycenter. For a detector at fixed position and orientation, 
at the SSB the continuous gravitational-wave signal is described by a 
sinusoid of constant amplitude and a phase given by
\begin{equation}
   \Psi(t_{\SSB}) = \Phi_0 + \Phi(t_{\SSB}) = \Phi_0 + 2\pi  \sum_{k=0}^{\infty} \, \frac{f^{(k)}}{(k+1)!} 
  \,  t_{\SSB}^{k+1},
\end{equation}
where $\Phi_0$ is the initial phase, $\fkdot{0} \equiv f$ denotes the 
frequency, and $\fkdot{k>0}$ the $k$'th frequency time-derivative, 
defining every parameter
$\fkdot{k}$ at $t=0$ at the SSB.  The integer~$s>0$ denotes the number
of frequency time-derivatives to be taken into account, therefore it is set
$\fkdot{k>s}=0$. In the case of an isolated rapidly rotating neutron
star with non-axisymmetric deformations and negligible proper motion~\cite{negprop}, 
the waveforms corresponding to the plus~($+$) and cross~($\times$) 
polarizations are 
\begin{equation}
  h_+ (t) = A_+ \, \sin \Psi(t), \qquad 
  h_\times (t) = A_\times \, \cos \Psi(t).
\end{equation}

The Earth's motion with respect to the SSB leads to
Doppler effects causing amplitude and phase modulations of 
the CW signal received at the detector.
Define $\vec n$ as the constant unit vector pointing from the SSB to the source.
Neglecting relativistic and higher order corrections, a wavefront
arriving at the detector at time $t$, passes the SSB at time
 \begin{equation}
   t_{\SSB} = t + \frac{\vec r (t) \cdot \vec n}{c} \;,
\end{equation}
where the vector $\vec r (t)$ connects from the SSB to the detector, 
and $c$ is the speed of light.

It is shown in~\cite{jks1} that the resulting phase evolution of the continuous 
gravitational-wave signal can be reproduced without significant loss
in signal-to-noise ratio by the model
\begin{equation}
   \Phi(t) = 2\pi\,\sum_{k=0}^{s} \left[ \frac{f^{(k)}\,t^{k+1}}{(k+1)!}
  + \frac{\vec{r}(t)}{c}\cdot \vec{n}\, \frac{f^{(k)}\,t^k}{k!}\right].
   \label{e:phase model}
\end{equation}
The received signal is also amplitude modulated by the varying
antenna-pattern of the detector due to its motion with the rotation of 
the Earth. The dimensionless signal response function~$h(t)$ of an 
interferometric detector  to a weak plane gravitational wave in the 
long-wavelength approximation  is
a linear combination of the form:
\begin{equation}
  h(t) = F_+ (t) \, h_+(t) + F_\times (t) \, h_\times(t) ,
  \label{e:h_t}
\end{equation}
where $F_{+,\times}$ are called the antenna-pattern functions,
resulting in the amplitude modulations from Earth's spinning motion.
They lie in the range $-1\le F_{+,\times}\le 1$,
and depend on the orientation of the detector and source, 
and on the polarization angle~$\psi$  of the waves.

The optimal detection statistic~\cite{{jks1},{lrr-2005-3}} obtained from the likelihood 
ratio~$\Lambda$ defines the matched filter
\begin{equation}
  \ln \Lambda = \frac{T}{S_h} \left[  (x||h) - \frac{1}{2} (h||h)\right],
  \label{e:loglikelihood}
\end{equation}
where $S_h$ is the one-sided noise strain spectral density which
is assumed to be constant over the narrow bandwidth of the signal, 
and the inner product is defined as
\begin{equation}
  (x||y) \equiv \frac{2}{T} \int_{-T/2}^{T/2}\, x(t)\,y(t)\, \rmd t ,
\end{equation}
centering the coherent observation-time interval of duration~$T$ around $t=0$.
Replacing the amplitude parameters in \Eref{e:loglikelihood} by their values
which maximize $\ln \Lambda$, the so-called maximum likelihood (ML) estimators,
defines the detection statistic~$\F$ as
\begin{equation}
  \F \equiv \ln \Lambda_{\rm ML}.
  \label{e:Fstatistic}
\end{equation}

\section{Matched-filtering detection statistic of a simplified signal model}
\label{sec:simple-matched-filtering}

\subsection{The simplified signal model}
\label{ssec:OrbitalPhase}

The phase of the continuous gravitational-wave signal is expected to change
very rapidly at the detector site on the Earth over a characteristic time length of 
typically less than ten seconds, whereas the amplitude of the signal changes 
with a period of one sidereal day.  As a result~\cite{jk2}, the detection of a CW 
signal requires an accurate model of its phase, because even $1/4$ of a cycle
difference between template and signal can lead to a loss in signal-to-noise
ratio of 10\%.  Whereas modeling of its amplitude is less critical. 
Therefore, the antenna pattern functions $F_{+,\times}$ are assumed to be constant, 
so that the signal model~(\ref{e:h_t}) becomes 
\begin{equation}
  h(t) = A_1 \cos \Phi(t) + A_2 \sin \Phi(t)  \;,
  \label{e:const-ampliudes}
\end{equation}
where $A_{1,2}$ are defined to be the constant effective amplitudes. The validity
of this approximation is investigated using Monte Carlo simulations in~\cite{jk2}.

The vector $\vec{r}(t)$ connecting the SSB and the detector can be  decomposed into
an orbital component $\vec{r}_{\rm orb}(t)$ and a spin component
$\vec{r}_{\rm spin} (t)$ as
\begin{equation}
  \vec r (t) = \vec r_{\rm orb} (t) + \vec r_{\rm spin} (t) \;,
  \label{e:det motion}
\end{equation}
where $\vec r_{\rm orb} (t)$ represents the vector from the SSB to the Earth's 
barycenter, and $\vec r_{\rm spin} (t)$ is the vector from the Earth's barycenter 
to the detector. Thus, substituting the decomposition~\eref{e:det motion} into~\Eref{e:phase model} 
one can write separately the orbital component~$\phi_{\rm orb}(t)$ and 
the spin component~$\phi_{\rm spin}(t)$ in the phase model~\eref{e:phase model} as
\begin{equation}
  \Phi(t) = 2\pi \left( \sum_{k=0}^{s} \frac{\fkdot{k}}{(k+1)!}\,t^{k+1}\right)  
  + \phi_{\rm orb}(t) + \phi_{\rm spin}(t) ,
  \label{e:phase model 2}
\end{equation}
where
\begin{eqnarray}
   \phi_{\rm orb}(t) &\equiv& 2\pi\,\frac{\vec r_{\rm orb} (t) \cdot \vec{n}}{c}  
   \left( \sum_{k=0}^{s} \frac{f^{(k)}}{k!}t^k\right),\label{e:phase-orb-part}\\ 
   \phi_{\rm spin}(t) &\equiv& 2\pi\,\frac{\vec r_{\rm spin} (t) \cdot \vec{n}}{c}  
   \left( \sum_{k=0}^{s} \frac{f^{(k)}}{k!}t^k\right)\label{e:phase-spin-part}.
\end{eqnarray}

The orbital motion of the Earth has a period of one year, so its
angular frequency is $\Omega_{\rm orb} = 2\pi/1\,{\rm yr}$.
Fully coherent all-sky searches for observation times $T$ much larger
than a  few days are computationally prohibitive~\cite{{jks1}, {bccs1:1998}}.
Thus for computationally feasible coherent searches or coherent stages of
hierarchical multistage searches~\cite{{hough:2005},{cutler:2005}} the typical observation
time baseline would be of order a few days.
Therefore, only observation times $T$ are considered, which are much 
shorter than one year: $\Omega_{\rm orb} T \ll 2\pi$. 
Then the component  
$\vec{r}_{\rm orb}(t)$ will vary slowly and one may use a Taylor expansion
at time $t_0$ with
\begin{equation}
  \vec{r}_{\rm orb}(t) = \sum_{\ell=0}^\infty \,
  {\vec{r}}^{\,(\ell)}_{\rm orb}(t_0) \, \frac{(t-t_0)^{\ell}}{\ell!} ,
  \label{e:r-orb}
\end{equation}
where $ {\vec{r}}^{\,(\ell)}_{\rm orb}(t_0)$ denotes the $\ell$'th derivative with respect to
time of $\vec{r}_{\rm orb}(t)$ evaluated at $t_0$. Without loss of generality we may 
choose $t_0=0$ as the midpoint of the observation of duration $T$ in 
the following discussion. Define $\vec{\xi} \equiv \vec{r}_{\rm orb}(0)/c$
with $\xi \equiv |\vec\xi|$ such that
\begin{equation}
  \xil{\ell} = \frac{{\vec{r}}^{\,(\ell)}_{\rm orb}(0)}{c} \qquad {\rm and}\qquad
  \xi^{(\ell)} = \left|  \xil{\ell}  \right| .
\end{equation}
Together with \Eref{e:r-orb} the orbital component of the phase~\eref{e:phase-orb-part}
can be written as
\begin{eqnarray}
  \phi_{\rm orb}(t) &=& 2\pi \left(  \sum_{k=0}^{s} \fkdot{k} \frac{t^k}{k!} \right)
  \left(  \sum_{\ell=0}^\infty \,
  \frac{t^{\ell}}{\ell!}  \, \xil{\ell}\cdot \vec{n} \right)  \nonumber \\
  &=& 2\pi \left[ \sum_{k=0}^\infty \,t^k \left(  \sum_{\ell=0}^k 
  \frac{\fkdot{\ell}\;\xil{k-\ell} \cdot \vec{n}}{\ell!\,(k-\ell)!} \right)  \right] .
  \label{e:orb-phase}
\end{eqnarray}

The spinning motion of the Earth has a period of one sidereal day ($1\,{\rm sd}$),
which translates into an angular frequency of $\Omega_{\rm spin}~=~2\pi/1\,{\rm sd}$.
The corresponding average velocity of $v_{\rm spin}/c~\approx~10^{-6}$ is two
orders of magnitude smaller than the corresponding orbital velocity,
$v_{\rm orb}/c~\approx~10^{-4}$.
In what follows we neglect the contribution of the spin component
$\phi_{\rm spin}(t)$ to the phase~\eref{e:phase model 2}.
\Sref{sec:Xspin} will discuss in detail the effects of the spin component
$\phi_{\rm spin}(t)$ in terms of the matched-filtering amplitude.

Using $\phi_{\rm orb}(t)$ in the form of \Eref{e:orb-phase}, we refer to
$\Phi_{\rm orb}(t)$ as the ``orbital phase model":
\begin{eqnarray}
   \Phi_{\rm orb}(t)  &\equiv& 2\pi \Biggl[ f \, \vec{\xi} \cdot \vec n
  + \sum_{k=0}^{\infty} \, t^{k+1}\Biggl( \frac{\fkdot{k}}{(k+1)!} \nonumber\\
  & &  + \sum_{\ell=0}^{k+1} \frac{\fkdot{\ell}}{\ell! (k-\ell+1)!} \, 
  \xil{k-\ell+1} \cdot \vec{n} \Biggr) \Biggr] .
  \label{e:orbital-phase}
\end{eqnarray}
By reparameterization the orbital phase model~\eref{e:orbital-phase} can be written as
\begin{equation}
  \Phi_{\rm orb}(t) = \sum_{m=0}^\infty \, u_m \, t^m ,
  \label{e:orbital-phase-short}
\end{equation}
where  the coefficients~$u_m$ of the power series are defined by
\begin{subequations}\label{e:u-all}
\begin{widetext}
\begin{eqnarray}
  u_0 &\equiv&  2\pi f \, \vec{\xi} \cdot \vec n , \\
  u_1 &\equiv& 2\pi \left( f + f \, \xil{1} \cdot \vec n
  + \fkdot{1}  \,\vec{\xi} \cdot \vec n  \right) ,\\
  u_2 &\equiv& 2\pi \left( \frac{\fkdot{1}}{2} + \frac{f}{2} \, \xil{2} \cdot \vec n 
  + \fkdot{1} \, \xil{1} \cdot \vec n +\frac{ \fkdot{2}}{2}\, \vec{\xi} \cdot \vec n \right) ,\\
   u_3 &\equiv& 2\pi \left( \frac{\fkdot{2}}{6} + \frac{f}{6} \, \xil{3} \cdot \vec n 
  + \frac{\fkdot{1}}{2} \, \xil{2} \cdot \vec n + \frac{\fkdot{2}}{2}\, \xil{1} \cdot \vec n 
  + \frac{\fkdot{3}}{6}\, \vec{\xi} \cdot \vec n\right) ,
\end{eqnarray}
\end{widetext}
so that for arbitrary order of $m > 0$ the coefficient $u_m$ is obtained as
\begin{equation}
  u_m \equiv 2\pi \left( \frac{\fkdot{m-1}}{m!} +  \sum_{\ell=0}^m 
  \frac{\fkdot{\ell}}{\ell! (m-\ell)!} \, \xil{m-\ell} \cdot \vec{n}  \right)  .
  \label{e:u-m}
\end{equation}
\end{subequations}

\subsection{The simplified matched-filtering detection statistic}
\label{ssec:SimpleMatchedFilter}

By analogy to \Eref{e:loglikelihood}, we refer to $\ln \Lambda^{\star}$
as the log likelihood function of the simplified CW signal model described
in previous Section~\ref{ssec:OrbitalPhase}.
Maximization of $\ln \Lambda^{\star} $ with respect to
the unknown amplitudes $A_{1,2}$ yields their ML estimators.
By substituting the ML estimators back into $\ln \Lambda^{\star} $, the 
simplified detection statistic $\Xstat$ is defined as
\begin{equation}
  \Xstat \equiv \ln \Lambda_{\rm ML}^{\star} 
  = \frac{T}{2\,S_h}\,\left| (x||\rme^{-i \Phi_{\rm orb}}) \right|^2
  = \frac{T}{2\,S_h}\, |\Xamp|^2 ,
  \label{e:Xstat}
\end{equation}
where the detection-statistic amplitude~$\Xamp$ has been defined through 
\begin{equation}
  \Xamp \equiv (x||\rme^{-i \Phi_{\rm orb}}) ,
\end{equation}
using the orbital phase model~$\Phi_{\rm orb}$.

For further simplicity consider a data set $x(t)$ which only contains unit-amplitude 
signal~$\textrm{s}(t)$, such that 
\begin{equation}
  x(t) = \Re[\textrm{s}(t)]. 
\end{equation}
Let the phase-parameter vector $\vDoppler_\sig=(\{f_\sig^{(k)}\},\vec{n}_\sig)$ define the 
phase $\Phi_{\rm orb}^\sig(t)$ of the signal. Then~$\textrm{s}(t)$ can be expressed as
\begin{equation}
  \textrm{s}(t) = \rme^{-i \Phi_{\rm orb}^\sig(t)} ,
  \label{e:signal}
\end{equation}
and one obtains $x(t)=\cos \Phi_{\rm orb}^\sig(t)$.
The difference in phase $\Delta \Phi_{\rm orb}(t)$ between the 
phase $\Phi_{\rm orb}^\sig(t)$ of the signal and the 
phase $\Phi_{\rm orb}(t)$ of a template~$\vDoppler=(\{f^{(k)}\},\vec{n})$ is defined  by
\begin{equation}
  \Delta \Phi_{\rm orb}(t) \equiv \Phi_{\rm orb}^\sig(t) - \Phi_{\rm orb}(t).
  \label{e:DeltaPhi}
\end{equation}

The maximization of $\Xstat$ is equivalent to maximizing~$|\Xamp|^2$.
Using \Eref{e:signal} one may rewrite the simplified matched-filtering 
amplitude~$\Xamp$  as
\begin{eqnarray}
  \Xamp &=& \frac{2}{T} \int_{-T/2}^{T/2} \, x(t) \, 
  \rme^{-i \Phi_{\rm orb}(t)} \, \rmd t \nonumber\\
  &=& \frac{1}{T} \int_{-T/2}^{T/2} \, \rme^{i \Delta \Phi_{\rm orb}(t)}
  + \rme^{-i \left[\Phi_{\rm orb}^\sig(t)+\Phi_{\rm orb}(t)\right]} \, \rmd t .
  \label{e:X}
\end{eqnarray}
Dropping the rapidly oscillating term in \Eref{e:X} yields
\begin{equation}
  \Xamp \approx \frac{1}{T} \int_{-T/2}^{T/2} \, \rme^{i \Delta \Phi_{\rm orb}(t)} \, \rmd t .
  \label{e:Xamp-appr}
\end{equation}
Thus, \Eref{e:Xamp-appr} shows that  $|\Xamp|$ has a global maximum of $|\Xamp|=1$,
if during the observation time interval~$T$ the phase 
difference~$\Delta \Phi_{\rm orb} (t)$ is stationary:
\begin{equation}
  \frac{\partial \Delta \Phi_{\rm orb} (t)}{\partial t} = 0. 
  \label{e:maxcond}
\end{equation}

Later, we will demonstrate that for the orbital 
phase model, unity of~$|\Xamp|$  can 
be obtained {\it not only} for the case where
all individual phase parameters exactly match 
($\Delta \fkdot{k} \equiv \fkdot{k}_\sig - \fkdot{k} =0$, 
$\Delta \vec n \equiv  \vec{n}_\sig - \vec n= \vec 0$) and so $\Delta \Phi_{\rm orb} (t)=0$, 
{\it but also} for different non-zero offsets 
($\Delta f^{(k)} \ne 0 , \Delta \vec n \ne \vec 0$)  which compensate 
each other to achieve $\partial \Delta \Phi_{\rm orb} (t) /\partial t \approx 0$ and thus
lead to a value of~$|\Xamp|$ close to unity.

\section{Global-correlation hypersurfaces in parameter space
\label{sec:Global correlations}}

\subsection{The global-correlation equations}
The central goal of this work is to identify those locations
in parameter space where the simplified detection statistic~$\Xstat$
is maximal, which corresponds to regions where $|\Xamp|^2$ is one.
Consistent with Equations~\eref{e:orbital-phase-short}, \eref{e:u-all} and~\eref{e:DeltaPhi}, 
the differences between the coefficients~$u_m$ of the template's phase and 
the coefficients~$u^\sig_{m}$ of the signal's phase are defined by
\begin{equation}
  \Delta u_m \equiv u^\sig_{m} - u_m \,.
\end{equation}
Thus, \Eref{e:DeltaPhi} can be expressed as
\begin{equation}
  \Delta \Phi_{\rm orb}(t) = \sum_{m=0}^{\infty} \Delta u_m\, t^m,
  \label{e:DPhiPolynom}
\end{equation}
and finally one can rewrite \Eref{e:X} as
\begin{equation}
  \Xamp 
  = \frac{\rme^{i \Delta u_0}}{T} \int_{-T/2}^{T/2} \,
  \exp \left(i \sum_{m=1}^\infty \Delta u_m \, t^m \right) \, \rmd t .
  \label{e:orb-X-2}
\end{equation}
It is apparent that $|\Xamp|$ does not depend upon the 
zero-order term $\Delta u_0$, and therefore the same holds for
the detection statistic~$\Xstat$.
The values of $\Delta u_m$ for which $|\Xamp|$ attains its maximum 
of $|\Xamp|=1$ consistent with~\eref{e:maxcond} are obvious from \Eref{e:DPhiPolynom},
namely when 
\begin{equation}
  \frac{\partial \Delta \Phi_{\rm orb}(t)}{\partial t} =  \sum_{m=0}^{\infty} m\, \Delta u_m\, t^{m-1} = 0,
\end{equation}  
the following central result is obtained: the family of {\it global-correlation equations}
which describes the global maximum structure of $\Xstat$,
\begin{equation}
  \Delta u_m = 0,
  \label{e:Dum0}
\end{equation}  
where $m > 0$, because $\Xstat$ is independent of $\Delta u_0$.
Because~$(1,t,t^2,\dots)$ is a basis of the vector space of real polynomials,
the zero vector can only be represented by the trivial linear combination, which
is given by \Eref{e:Dum0}. 

Thus, the first global-correlation equation given by $\Delta u_1 = 0$ 
can be rewritten as
\begin{equation}
   f + f \, \xil{1} \cdot \vec n + \fkdot{1}  \, \vec{\xi} \cdot \vec n = K_1 ,
  \label{e:GCE1}
\end{equation}
where $K_1 \equiv  u^\sig_{1} /2\pi$ represents a constant  defined by  the 
signal's phase parameters.
As a side remark, it should be mentioned that \Eref{e:GCE1} is 
a generalization of the first-order global-correlation equation 
found in~\cite{prixitoh:2005} to signals with non-zero frequency time-derivatives.
But in order to describe qualitatively the global maximum structure of the simplified 
detection statistic in parameter space, considering only the first order 
(as done in~\cite{prixitoh:2005}) might not be sufficient. As a matter of fact,
this is shown in \Sref{ssec:third-order-est} and will be confirmed by 
analyzing simulated signals later on.

Therefore, continuing to the next order in time, one obtains from the 
second condition $\Delta u_2 = 0$ the following relation, 
\begin{equation}  
  \fkdot{1} + f \,\xil{2} \cdot \vec n 
  + 2 \fkdot{1}  \,\xil{1} \cdot \vec n 
  + \fkdot{2} \,\vec{\xi} \cdot \vec n= K_2 ,
  \label{e:GCE2}
\end{equation}
where the constant $K_2 \equiv  u^\sig_{2} /2\pi$ is defined by the signal's phase parameters.

In general, this scheme can extended to arbitrary order $m > 0$. Thus, 
the family of global-correlation equations represented by
$\Delta u_m = 0$ can be written in the form
\begin{equation}
  \frac{\fkdot{m-1}}{m!} + \sum_{\ell=0}^m 
  \frac{\fkdot{\ell}\,\xil{m-\ell} \cdot \vec{n} }{\ell! (m-\ell)!}  = K_m ,
  \label{e:m-fam}
\end{equation}
where  the signal's phase parameters determine the 
constant $K_m \equiv  u^\sig_{m} /2\pi$.
In the following the geometry of the solution to the family 
of equations given by~(\ref{e:m-fam}) will be explored and the {\it hypersurfaces} 
they describe in parameter space will be illustrated.

\subsection{The geometry of the global-correlation equations}
\label{ssec:geometry}

Let the space of phase parameters~$\vDoppler=(\{\fkdot{k}\},\vec n)$ be a 
manifold denoted by~$\Pspace$.
Previously we defined $\vec n$ as a unit vector pointing to the 
source's sky location in the SSB frame of reference. A position on the sky
can be determined by two independent coordinates, for example one can
use equatorial coordinates of right ascension (RA) and
declination, denoted by~$\alpha$ and~$\delta$, respectively. In these coordinates:  
$\vec n = (\cos \delta \, \cos \alpha, \cos \delta \, \sin \alpha, \sin \delta)$. 
Recall that the integer $s$ is related to  the number of spin-down parameters 
considered ($\fkdot{k>s} = 0$), thus the phase-parameter space 
dimensionality is $s+3$.

By inspection of \Eref{e:m-fam}, it is obvious that for a given continuous 
gravitational-wave signal the set of solutions to each global-correlation equation
$\Delta u_m=0$, is a {\it hypersurface} in the search parameter space~$\Pspace$.
Denoting each hypersurface by~$ \Hyps_m$ one may write
\begin{equation}
   \Hyps_m = \{ \vDoppler \in \Pspace \; :\; \Delta u_m = 0\},
   \label{e:Hypersurface}
\end{equation}
given the signal's phase parameters $u^\sig_{m}$.
The dimensionality of each hypersurface~$\Hyps_m$ is $s+2$.

Using the parameters~$u_m$ of \Eref{e:u-all} define the
vector~$\vecu\equiv(\{u_m\})=(u_1,u_2,\dots)$ and denote the corresponding
parameter-space manifold by~$\Uspace$. 
Let the vector~$\vecu^{\sig} \in \Uspace$ denote the signal's parameters,
and define the difference $\Delta \vecu = \vecu^\sig - \vecu = (\Delta u_1, \Delta u_2, \dots)$.
Considering the detection-statistic amplitude~$|\Xamp|$  as a function 
of $\Delta \vecu$, then this function is extremal with
respect to the parameter~$u_m$ along the hypersurface~$\Hyps_m$ 
defined by $\Delta u_m=0$,
\begin{equation}
 \frac{\partial\, \left|\Xamp (\Delta \vecu)\right|}{\partial u_m}\biggr|_{\Delta u_m = 0} = 0.
\end{equation}
Obviously, on the intersection of these hypersurfaces~$\Hyps_m$ at $\Delta \vecu = 0$, 
$|\Xamp|$ is maximum with respect to all the parameters~$u_m$:
$\nabla \left|\Xamp(\Delta \vecu=0)\right| =0$.
Thus, it follows that the simplified detection statistic~$\Xstat$
is also expected to be maximal at the intersection of these hypersurfaces. 
Therefore, the global maximum structure is defined by the intersection
of the global-correlation hypersurfaces~$\Hyps_m$.
This idea has been shown schematically in Figure~\ref{f:intersectdemo}.
In the absence of noise, all candidate events produced by a 
given CW signal which follows the simplified model introduced 
in \Sref{ssec:OrbitalPhase} will be located 
on the hypersurface~$\Hyps_1$, described by \Eref{e:GCE1}.
From all candidate events located on~$\Hyps_1$ those will belong to 
loudest ones (have largest values of~$\Xstat$) which are located 
at the intersection with hypersurface~$\Hyps_2$, 
which is described by \Eref{e:GCE2}. For each higher order~$m$ 
this behavior carries itself forward in the same way.

\subsection{A visualizing example of the global-correlation hypersurfaces}
\label{ssec:example}

As an illustrative example visualization of  the geometrical structure 
formed by the global-correlation hypersurfaces~\eref{e:m-fam}, 
we choose the four-dimensional phase-parameter space ($s=1$).
Thus, the four dimensions are sky position
(right ascension~$\alpha$ and declination~$\delta$), 
the frequency~$f$, and the first spin-down parameter~$\fkdot{1}$. 
For demonstration purposes, consider an exemplary continuous 
gravitational-wave signal with the phase parameters 
$f_\sig = 100.0\,\Hz$, $\alpha_\sig=2.0\,{\rm rad}$, $\delta_\sig=-1.0\,{\rm rad}$  and
$\fkdot{1}_\sig = -10^{-10}\,\Hz / {\rm s}$. 

In this illustration, assume that we have a bank of templates
available covering the whole sky, the entire Doppler frequency band~\cite{earthdoppler},
$f_\sig\pm f_\sig\times10^{-4}$, and for the spin-down value 
of  $\fkdot{1} =-10^{-11}\,\Hz/{\rm s}$.
The attentive reader will notice that the signal's spin-down value is not covered 
by this template value, but mismatched by an order of magnitude. 
This choice will demonstrate that due to the parameter-space correlations 
the signal is still expected to be detected in such a search, 
producing a detection-statistic maximum-structure which 
is predicted by the global-correlation equations.

The global-correlation equations only for $m \le 2$ have to be considered in the 
present example, because contributions to $|\Xamp|$ in \eref{e:orb-X-2} 
from terms beyond second order
are negligible, as will be shown later in \Sref{ssec:third-order-est}. 
There, the neglected third-order term in \Eref{e:orb-X-2} is estimated
and found to be irrelevant, in the case of the signal investigated here,
and for observation times $T$ approximately less than $61\,{\rm hours}$. 

For the case of $m \le 2$ the integral~\eref{e:orb-X-2} can be 
obtained analytically as follows
\begin{widetext}
\begin{eqnarray}
  \Xamp &\approx& \frac{\rme^{i \Delta u_0}}{T} \int_{-T/2}^{T/2} \,
  \rme^{i(\Delta u_1 \, t+\Delta u_2 \, t^2)} \, \rmd t \nonumber\\
  &=&\frac{\sqrt{\pi}}{2T\sqrt{\Delta u_2}}\, 
  \exp\left[i\left( \Delta u_0- \frac{{\Delta u_1}^2}{4\Delta u_2} + \frac{\pi}{4} \right)\right] 
  \, \left\{ \erf\left[\frac{(\Delta u_1+\Delta u_2 T)\rme^{-i\pi/4}}{2\sqrt{\Delta u_2}}\right]
  - \erf\left[\frac{(\Delta u_1-\Delta u_2 T)\rme^{-i\pi/4}}{2\sqrt{\Delta u_2}}\right]\right\},
  \label{e:X-m-le2}
\end{eqnarray}
\end{widetext}
where the error function $\erf(x)$ is defined by
\begin{equation}
  \erf(x) \equiv \frac{2}{\sqrt{\pi}} \, \int_0^x \rme^{-t^2} \rmd t .
\end{equation}

\begin{figure}[!htb]
	\includegraphics[width=0.999\columnwidth]
	{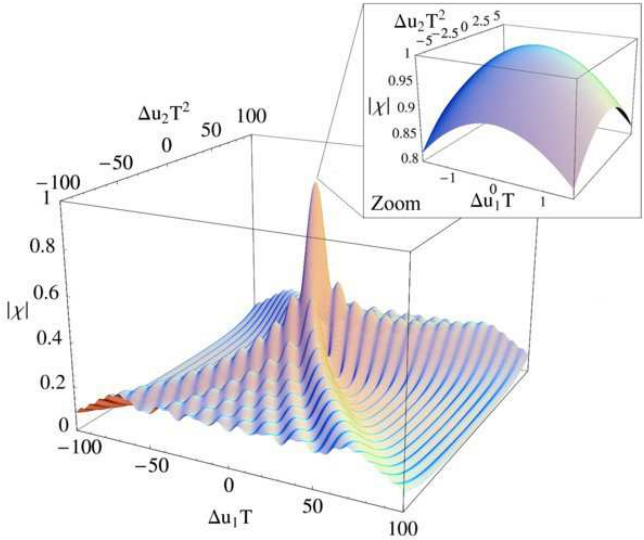}
	\caption{(Color online) Illustration of the simplified detection-statistic amplitude~$|\Xamp|$ for $m\le2$,
	given by \Eref{e:X-m-le2}, as a function of the dimensionless parameters 
	$\Delta u_1 T$ and $\Delta u_2 T^2$. 
	\label{f:X12}}
\end{figure}

\Fref{f:X12} shows $|\Xamp|$ given by \Eref{e:X-m-le2} as a 
function of the dimensionless parameters  $\Delta u_1 T$ and $\Delta u_2 T^2$.
This figure visualizes the fact that when one moves away in parameter space 
from the global-correlation hypersurfaces at $\Delta u_m=0$ to increasing non-zero
values of $\Delta u_m$, the detection-statistic amplitude~$|\Xamp|$ decreases rapidly 
from its maximum of one towards zero, as is shown in \Fref{f:X12}.

As mentioned earlier, the two equations from the family of global-correlation 
equations~(\ref{e:m-fam}) to be considered for the example signal studied here are:
\begin{subequations}\label{e:C12}
\begin{eqnarray}
  f + f \,\xil{1} \cdot \vec n  + \fkdot{1}  \,\vec{\xi} \cdot \vec n &=& K_1 ,
  \label{e:C1}\\
  \fkdot{1} + f \,\xil{2} \cdot \vec n + 2 \fkdot{1} \,\xil{1} \cdot \vec n &=&  K_2 
  \label{e:C2} ,
\end{eqnarray}
\end{subequations}
where in this case the constants $K_{1,2}$ are obtained from the signal's 
phase parameters~$(f_\sig,\fkdot{1}_\sig, \vec{n}_\sig)$ as
\begin{subequations}\label{e:K12}
\begin{eqnarray}
  K_1 &=& f_\sig + f_\sig \, \xil{1} \cdot \vec{n}_\sig 
  + \fkdot{1}_\sig \,\vec{\xi} \cdot \vec{n}_\sig ,
  \label{e:K1}\\ 
  K_2 &=& \fkdot{1}_\sig + f_\sig \, \xil{2} \cdot \vec{n}_\sig 
  + 2 \fkdot{1}_\sig \, \xil{1} \cdot \vec{n}_\sig \label{e:K2}.
\end{eqnarray}
\end{subequations}

\begin{figure*}
	\subfigure[
	Global-correlation hypersurface $\Hyps_1$
	and contour lines of constant frequency shown in the sky.
	\label{f:dsurfs1}]
	{\includegraphics[width=0.9\columnwidth]
	{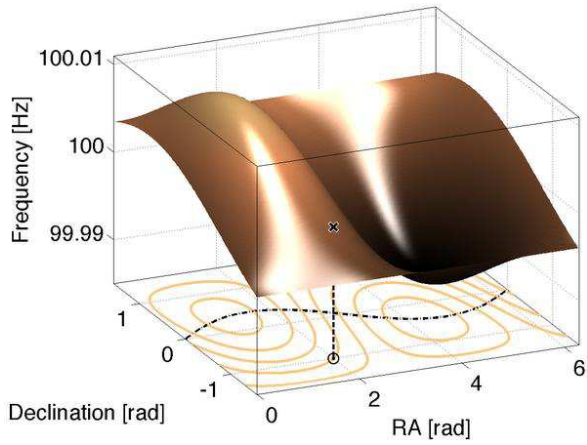}}\qquad
	\subfigure[
	Global-correlation hypersurface for $\Hyps_2$
	and contour lines of constant frequency shown in the sky (all close to each other).
	\label{f:dsurfs2}]
	{\includegraphics[width=0.9\columnwidth]
	{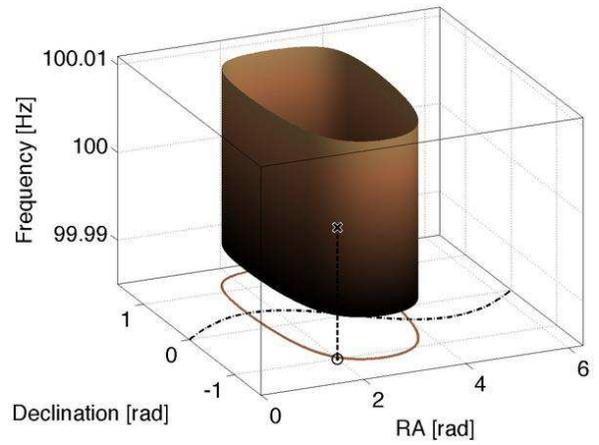}}\\
	\subfigure[
	Superposition of global-correlation hypersurfaces 
	$\Hyps_{1,2}$, and their frequency	contour lines in the sky.
	\label{f:dsurfs3}]
	{\includegraphics[width=0.9\columnwidth]
	{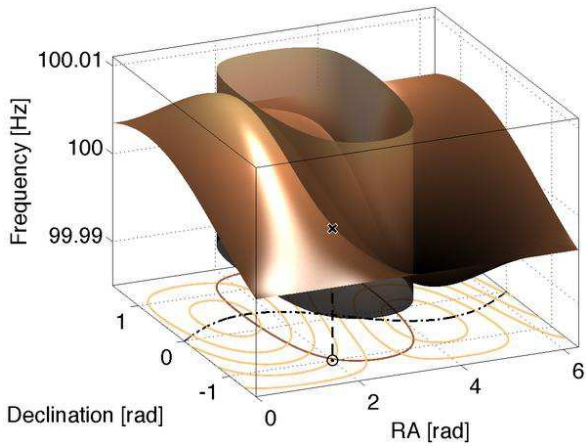}}\qquad \qquad
	\subfigure[
	Unit sphere with contour lines (circles) 
	of global-correlation hypersurfaces	$\Hyps_{1,2}$. 
	\label{f:dsurfs4}]
	{\includegraphics[width=0.7\columnwidth]
	{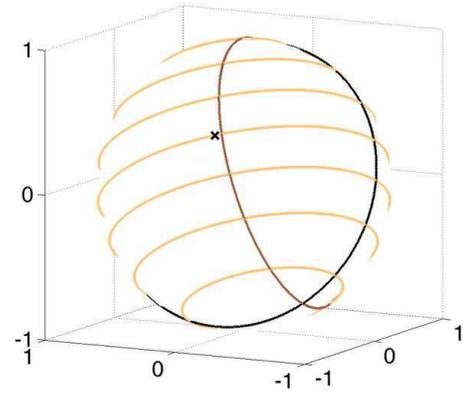}}\\
	\subfigure[
	Hammer-Aitoff sky projection showing the
	contour lines (circles) of global-correlation hypersurfaces $\Hyps_{1,2}$. 
	\label{f:dsurfs5}]
	{\includegraphics[width=\columnwidth]
	{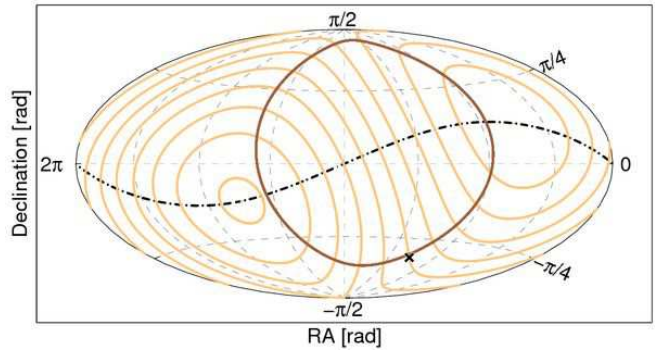}}
	\caption{
	(Color online) The global-correlation hypersurfaces $\Hyps_{1}$ ($\Delta u_1=0$) and 
	$\Hyps_{2}$ ($\Delta u_2=0$) for a given CW signal shown in the three-dimensional 
	subspace $\{f,\vec n \}$ at the fixed target spin-down of $\fkdot{1} =-10^{-11}\,\Hz/{\rm s}$.
	Each plot refers to GPS time~$793555944\,{\rm s}$.
	The phase parameters of the signal are $\alpha_\sig=2.0\,{\rm rad}$, 
	$\delta_\sig=-1.0\,{\rm rad}$, $f_\sig = 100.0\,\Hz$, 	$\fkdot{1}_\sig = -10^{-10} \Hz / {\rm s}$,
	as indicated by the black cross (black circle) in the subspace $\{f,\vec n \}$ (in the sky plane). 
	In~\subref{f:dsurfs3} the superposition of~\subref{f:dsurfs1} and~\subref{f:dsurfs2} illustrates 
         the locations of expected maximum detection statistic along the intersection curve
	of both hypersurfaces. In each plot the light solid contour lines in the sky of 
	constant~$f$ (and $\fkdot{1}$) are of $\Hyps_{1}$, the dark solid contour lines 
	are the ones of $\Hyps_{2}$.
	These contours of both hypersurfaces, shown in a Hammer-Aitoff projection in~\subref{f:dsurfs5},
	actually describe circles on the three-dimensional unit sphere as is
	visualized by~\subref{f:dsurfs4}. In the sky subspace of this example, the intersection curve
	approximately coincides with the contours of~$\Hyps_{2}$, and due to the mismatch in 
	spin-down the intersection curve does not pass through the signal's true sky-location.
	The black dotted-dashed curve represents the ecliptic.	
	\label{f:GCHypersurfaces} }
\end{figure*}

\Fref{f:GCHypersurfaces} presents the visualization of the hypersurfaces
$\Hyps_1$ and $\Hyps_2$ described by Equations~(\ref{e:C1})
and~(\ref{e:C2}) for the given CW signal. As described in Section~\ref{ssec:geometry}, when
choosing $s=1$, then we have $\dim \Hyps_1 = \dim \Hyps_2 = 3$.

In \Fref{f:GCHypersurfaces}  the subspace~$\{f,\vec n \}$ is shown at the fixed 
target spin-down of $\fkdot{1} =-10^{-11}\,\Hz/{\rm s}$.
The hypersurface described by \Eref{e:C1} for the fixed~$\fkdot{1}$ 
is depicted three-dimensionally in \Fref{f:dsurfs1}.
The 3D plot of \Fref{f:dsurfs2} shows the hypersurface defined 
by \Eref{e:C2} for the same signal and same template values. 
Finally, \Fref{f:dsurfs3} combines Figures~\ref{f:dsurfs1} and~\ref{f:dsurfs2} 
showing both hypersurfaces and illustrating their intersection. 
Along this intersection curve of both hypersurfaces, $|\Xamp|$ (and so~$\Xstat$) 
is expected to be maximal for the CW signal examined in this example.
The contour lines of constant~$f$ (and $\fkdot{1}$) in the sky 
of both hypersurfaces are shown on the three-dimensional unit-sphere
by \Fref{f:dsurfs4} and in a 2D Hammer-Aitoff projection by \Fref{f:dsurfs5}.
Here, in the sky subspace the intersection curve of $\Hyps_{1}$ and
$\Hyps_{2}$ (corresponding to maximal detection statistic) 
approximately coincides with the contours of $\Hyps_{2}$.

\subsection{Validity estimation of the global-correlation hypersurface description}
\label{ssec:third-order-est}

This section discusses how many hypersurfaces have to be considered
in order to describe the detection-statistic maximum-structure. In other words,
the question is investigated, in which region of the search parameter space and 
at which order~$m$ the power series~\eref{e:DPhiPolynom}
can be truncated to approximate the matched-filtering amplitude~$|\Xamp|$
better than a certain fractional loss one tolerates.

For a given order~$m$, one can estimate the contributions from the next-order 
term in the matched-filtering amplitude~$|\Xamp|$. As was done earlier in \Sref{ssec:example},
here we consider again the four-dimensional parameter space~$\{f,\fkdot{1}, \vec n \}$ of 
``typical" wide-band all-sky CW searches, with $f\le1\,{\rm kHz}$, $|\fkdot{1}| < f/\tau_{\rm min}$,
and a minimum spin-down age of $\tau_{\rm min}=50\,{\rm yrs}$.

\begin{figure}
	\centering
	\subfigure[\label{f:Xest1}]{\includegraphics[width=0.48\columnwidth]
	{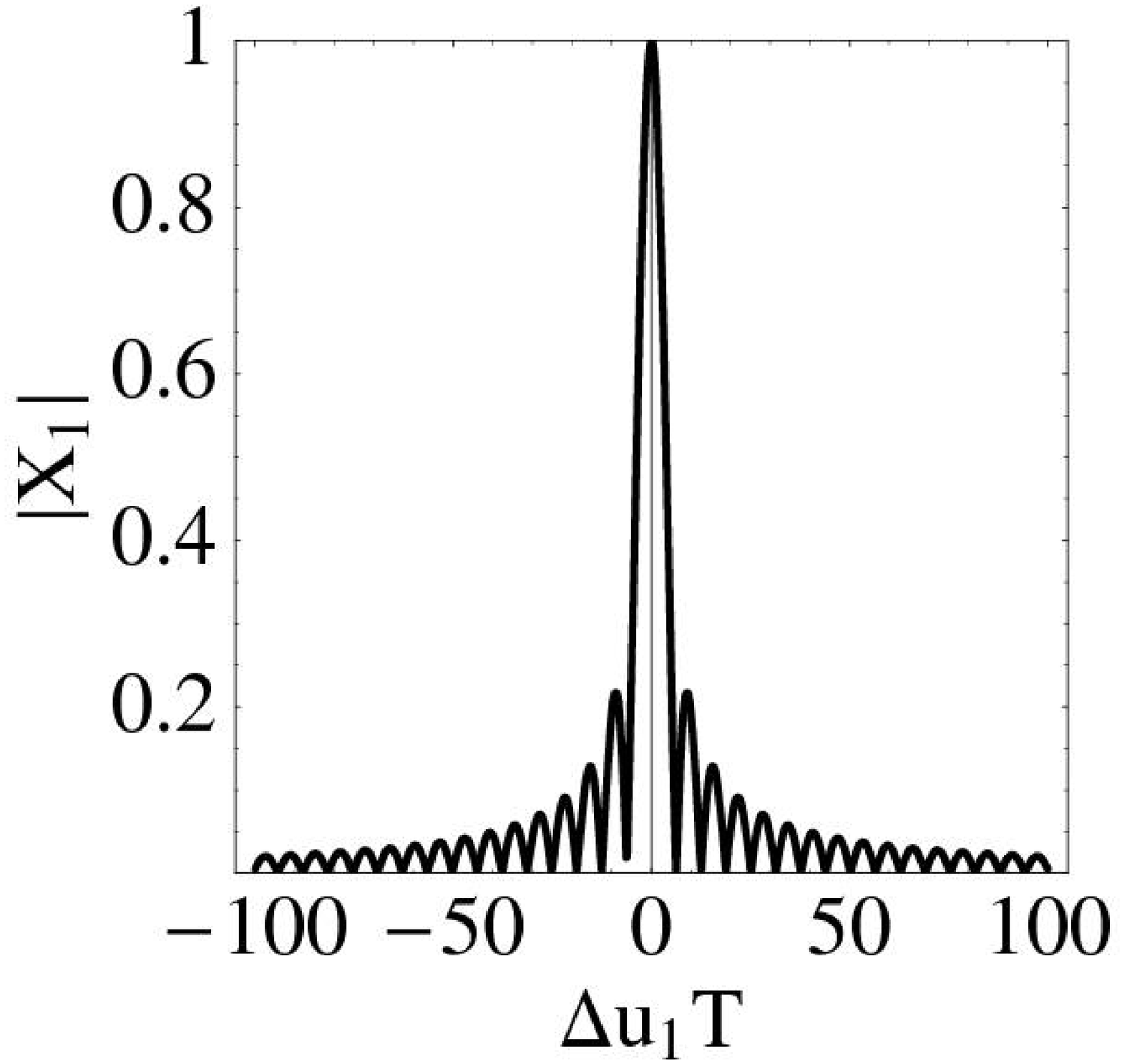}}\;\;
	\subfigure[\label{f:Xest2}]{\includegraphics[width=0.48\columnwidth]
	{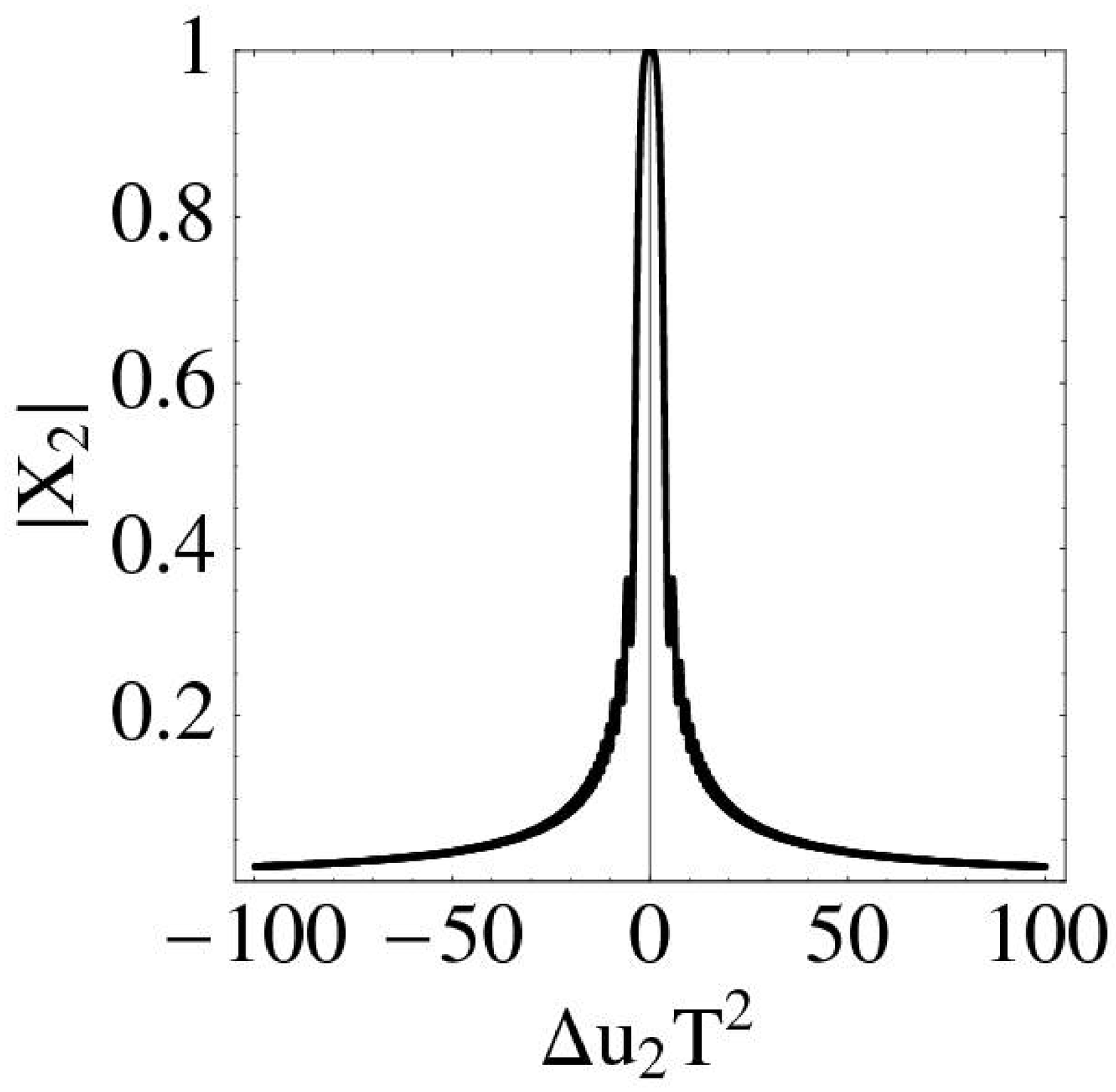}}\\
	\subfigure[\label{f:Xest3}]{\includegraphics[width=0.48\columnwidth]
	{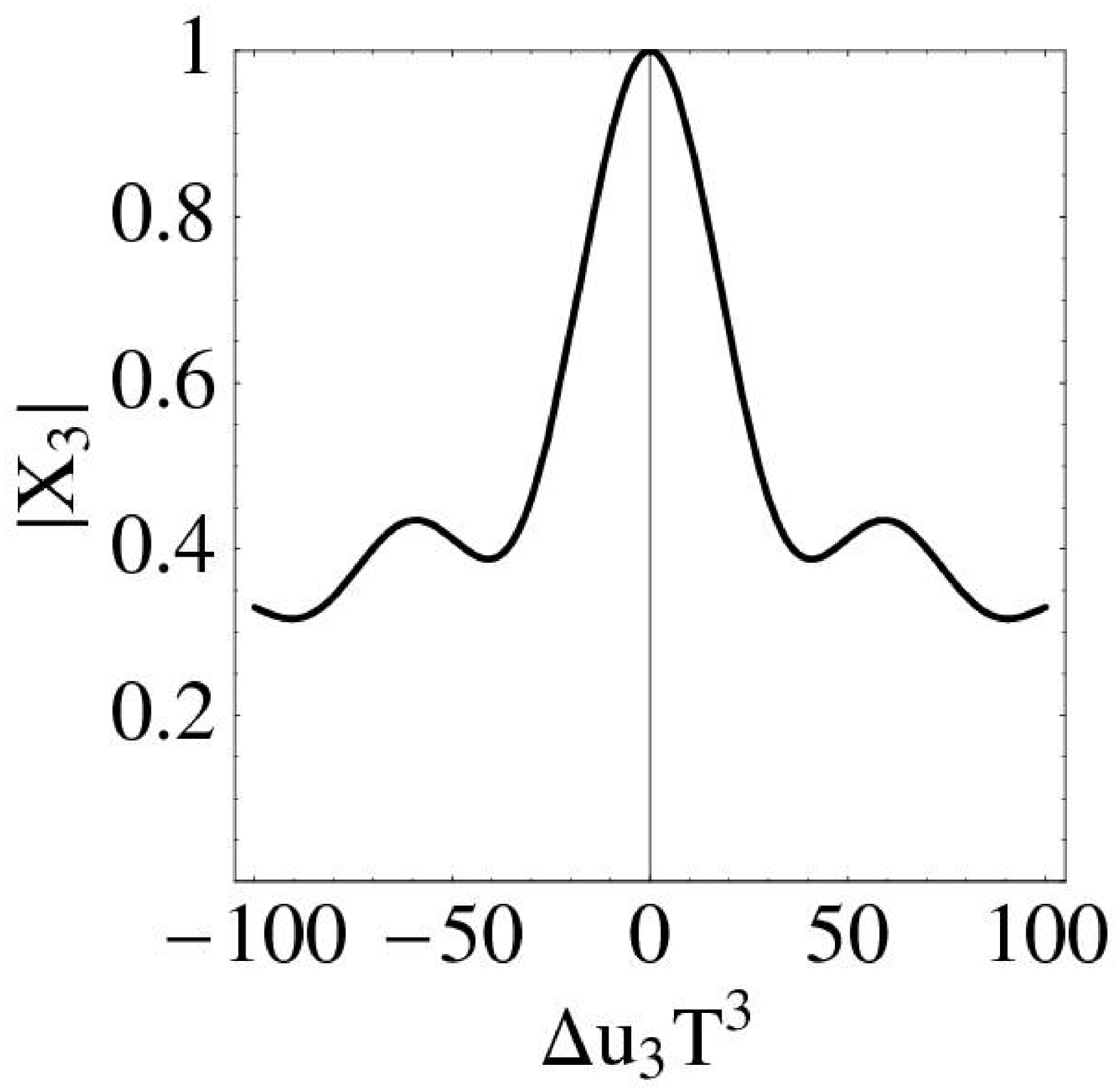}}\;\;
	\subfigure[\label{f:Xest4}]{\includegraphics[width=0.48\columnwidth]
	{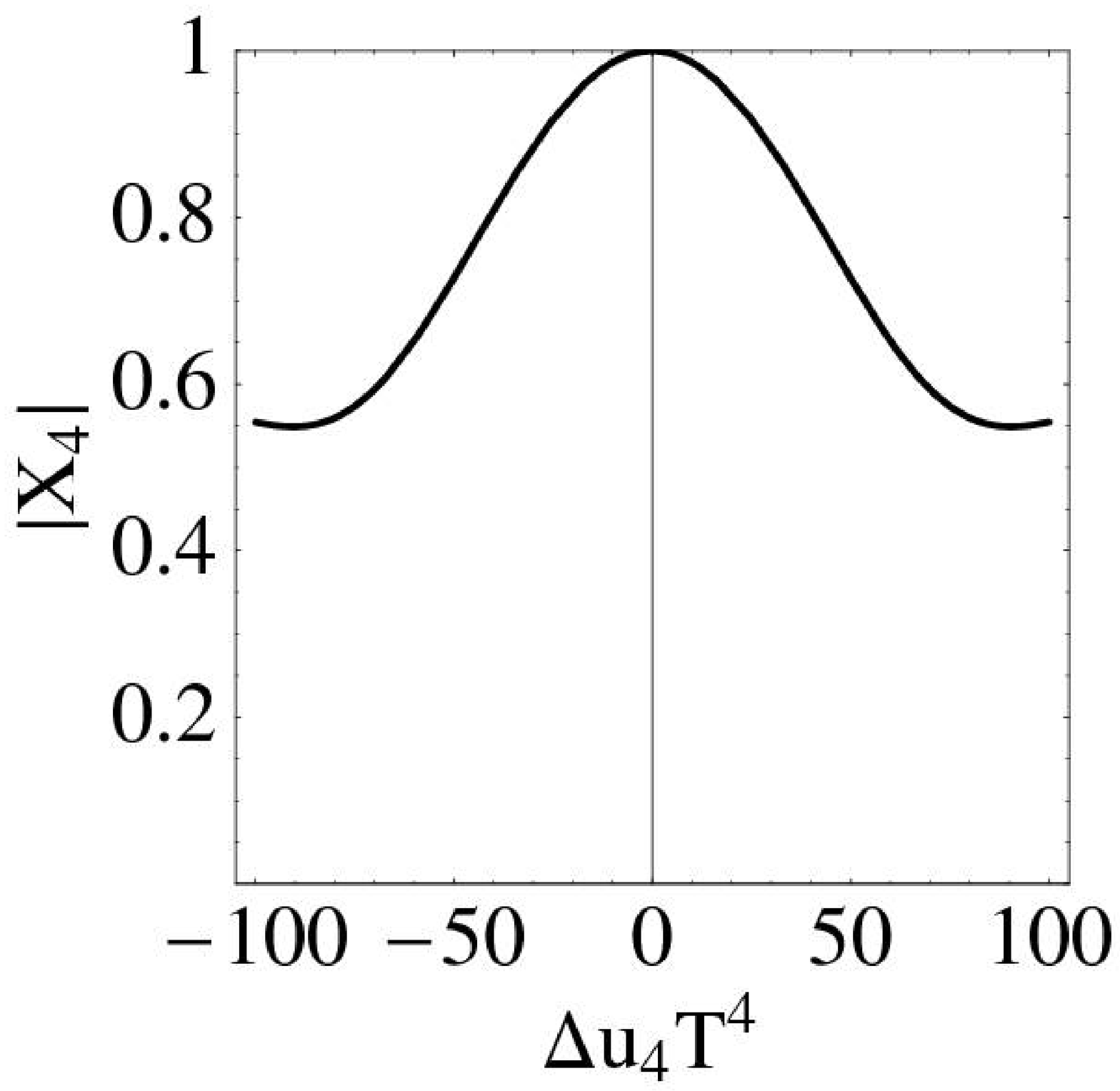}}
	\caption{Estimating the number of global-correlation hypersurfaces which contribute 
	significantly to the detection statistic. In each plot both axes are dimensionless.
	Further details are given in the text.
	\label{f:Xest}}
\end{figure}

To investigate the contribution from the first-order term in \Eref{e:orb-X-2} to~$|\Xamp|$, 
we compute the following integral ignoring terms in~$t$ with order $m>1$, 
\begin{equation}
  \Xest_1 \equiv \frac{1}{T}\int_{-T/2}^{T/2} \rme^{i\Delta u_1\, t} \rmd t 
  = \frac{2\sin(\Delta u_1\,T/2)}{\Delta u_1\,T}.
  \label{e:Xest1}
\end{equation}
 \Fref{f:Xest1} shows $|\Xest_1|$ as a function of 
the dimensionless parameter~$\Delta u_1\,T$.
As already explained earlier, $|\Xest_1|$ attains its global maximum of one, when
$\Delta u_1=0$, which is the first order global-correlation equation. 
If one requires for instance $|\Xest_1| \ge 0.9$, using \Eref{e:Xest1}
this yields the condition $|\Delta u_1|\,T \le 1.57$.
When expressing $\Delta u_1$ in terms of the original phase 
parameters using~\eref{e:u-all}, one obtains
\begin{eqnarray}
  |\Delta u_1| &\lesssim& 2\pi \Bigl[\, |\Delta f| \left(1+ \xia{1}\right) + 2f_\sig\, \xia{1} \nonumber\\
  && + |\Delta \fkdot{1} |\xi + 2 |\fkdot{1}_\sig|\,\xi  \,\Bigr].
  \label{e:Du1-Xest1}
\end{eqnarray}
But for typical wide-band all-sky CW searches with coherent observation times~$T$ of a 
day or a few days,  according to \Eref{e:Du1-Xest1} $|\Delta u_1|\,T$ can be
considerably larger than $1.57$.
Therefore, it is clear that the first-order term in \Eref{e:orb-X-2}  contributes significantly 
to the matched-filtering amplitude~$|\Xamp|$ and obviously cannot be neglected.

Similarly, we estimate the contribution from the second-order term given
that $\Delta u_1=0$ and by ignoring terms with $m>2$ in \Eref{e:orb-X-2}. 
We calculate the corresponding integral denoted by $\Xest_2$ as
\begin{eqnarray}
  \Xest_2 &\equiv& \frac{1}{T}\int_{-T/2}^{T/2} \rme^{i\Delta u_2\, t^2} \rmd t \nonumber\\
  &=& \frac{\rme^{i\pi/4}\sqrt{\pi}}{\sqrt{\Delta u_2\,T^2}}\,
  \erf\left[ \frac{\rme^{-i\pi/4}}{2}\sqrt{\Delta u_2\,T^2}\right].
  \label{e:Xest2}
\end{eqnarray}
In \Fref{f:Xest2}, $|\Xest_2|$ is shown as a function of $\Delta u_2\,T^2$. 
Obviously, $|\Xest_2|$ attains the maximum of one at the second order global-correlation
hypersurface, which is described by $\Delta u_2=0$. 
One finds that $|\Xest_2| \ge 0.9$ as long as $|\Delta u_2|\,T^2 \le 6.11$.
Reexpressing~$\Delta u_2$ in the original phase parameters yields
\begin{equation}
   |\Delta u_2| \lesssim 2\pi \left[  f_\sig \xia{2} 
   +  |\Delta \fkdot{1}| \left(\frac{1}{2}+\xia{1}\right) + 2|\fkdot{1}_\sig|\xia{1} \right].
   \label{e:Du2-Xest2}
\end{equation}
But according to \Eref{e:Du2-Xest2} for typical wideband all-sky CW searches 
with coherent observation times~$T$ of a day or a few days, $|\Delta u_2|\,T$
can take values considerably larger than $6.11$. 
Therefore, the second-order term in  \Eref{e:orb-X-2}  also contributes significantly 
to the matched-filtering amplitude~$|\Xamp|$ and cannot be neglected.

Continuing this scheme to the next order in~$t$, the contribution
from the third-order term in  \Eref{e:orb-X-2} is considered.
Given $\Delta u_1=0$ and $\Delta u_2=0$, we estimate
the contribution from the third-order term (with $m=3$) by evaluating:
\begin{eqnarray}
  \Xest_3 &\equiv& \frac{1}{T} \int_{-T/2}^{T/2}\,\rme^{i\,\Delta u_3 \, t^3} \rmd t \nonumber\\
  &=& \frac{\rme^{i\pi/6}}{3\left(\Delta u_3 T^3\right)^{1/3}}\,\biggl[2\,\Gamma\left(\frac{1}{3}\right) 
  - \Gamma\left(\frac{1}{3},\frac{i}{8}\,\Delta u_3\,T^3\right) \nonumber\\
  & & - \Gamma\left(\frac{1}{3},-\frac{i}{8}\,\Delta u_3\,T^3\,\right)\biggr] 
  \label{e:X3-b} ,
\end{eqnarray}
where the gamma function $\Gamma(x)$ and the upper incomplete gamma function
$\Gamma(a,x)$ are defined by
\begin{eqnarray}
  \Gamma(x) &\equiv& \int_{0}^{\infty}\, t^{x-1} \, \rme^{-t}\, \rmd t ,
  \label{e:gamma}\\
  \Gamma(a,x) &\equiv& \int_{x}^{\infty}\, t^{a-1} \, \rme^{-t} \, \rmd t .
  \label{e:upper-incgamma}
\end{eqnarray}
As can also be seen from \Fref{f:Xest3}, where $\Delta u_3\,T^3$ is plotted against $|\Xest_3|$, 
requiring $|\Xest_3| \ge 0.9$ leads to the condition
$|\Delta u_3|\,T^3 \le 9.79$. By rewriting $\Delta u_3$ using again the original phase 
parameters according to~\eref{e:u-all}, one obtains 
\begin{equation}
  |\Delta u_3| \lesssim 2\pi \left( \,\frac{ f_\sig}{3}\,\xia{3} +|\fkdot{1}_\sig|\xia{2}\,\right).
  \label{e:Du3-X3}
\end{equation}
By approximating 
\begin{equation}
  \xia{3} \approx \Omega_{\rm orb}^2\, \frac{v_{\rm orb}}{c}  \qquad {\rm and} \qquad
  \xia{2} \approx \Omega_{\rm orb}\, \frac{v_{\rm orb}}{c} ,
\end{equation} 
a condition is obtained regarding the observation time $T$ for which the 
contribution to $|\Xamp|$ from the third-order term in \Eref{e:orb-X-2} is negligible:
\begin{equation}
  T \lesssim \left[\frac{9.79 \,c}{2\pi\, \Omega_{\rm orb}\, v_{\rm orb}} \left(
  \frac{f_\sig}{3}\, \Omega_{\rm orb} 
  +|\fkdot{1}_\sig| \right)^{-1}\right]^{1/3}.
  \label{e:T-X3}
\end{equation}
In the example presented in Section~\ref{ssec:example}, a signal is considered with
$f_\sig =100.0\,\Hz$ and $\fkdot{1}_\sig=10^{-10}\,\Hz/{\rm s}$.
In this case, according to \Eref{e:T-X3} the third order can be neglected for
observation times $T \lesssim 61\,{\rm h}$. For a similar signal and search, 
but for instance at $f_\sig =1000.0\,\Hz$ the third-order term is
expected to be non-significant only for observation times $T \lesssim 28\,{\rm h}$.

However, considering cases where the third order in~$t$
is non-negligible, an analogous estimation of the contribution from the fourth-order term 
is necessary. Therefore, we calculate $\Xest_4$  as
\begin{eqnarray}
  \Xest_4 &\equiv&  \frac{1}{T} \int_{-T/2}^{T/2}\,\rme^{i\,\Delta u_4 \, t^4} \rmd t \nonumber\\
  &=& \frac{2}{\left( -i\,\Delta u_4\, T^4\right)^{1/4}} \nonumber\\
  & &\times \left[ 4\,\Gamma \left(\frac{5}{4}\right)
  -\Gamma \left(\frac{1}{4},-\frac{i}{16}\,\Delta u_4\, T^4\right) \right].
\end{eqnarray}
\Fref{f:Xest4} shows $|\Xest_4|$ as a function of $\Delta u_4\,T^4$.
From the condition $|\Xest_4| \ge 0.9$ follows $|\Delta u_4|\,T^4 \le 27.78$.
Using this, analogously to Equations~\eref{e:Du3-X3} and~\eref{e:T-X3}, an approximate
condition upon $T$ can be found, which in the case of the present example signal yields
that $|\Xest_4| \ge 0.9$ for $T$ being approximately less than $10\,{\rm days}$.
As for the wide-band all-sky CW searches described earlier,
coherent observation times are typically only of one day or a few days,
this means that in practice contributions from the fourth-order term in  \Eref{e:orb-X-2}
 to~$|\Xamp|$ are insignificant in such cases.

\section{Predictions by the global-correlation equations versus full $\F$-statistic}
\label{sec:prediction-vs-Fstat}
In this section the analytical predictions of the global-correlation 
equations~\eref{e:m-fam} based on the simplified detection 
statistic~$\Xstat$ are compared to the results of fully coherent searches 
using the detection statistic~$\F$ in different data sets 
containing artificial CW signals. 
The computation of the full $\F$-statistic includes 
effects of amplitude modulation and involves precise calculation 
of the detector position with respect to the SSB using an accurate
ephemeris model.

\subsection{Comparison with the $\F$-statistic for software-simulated signals
with no detector noise}
\label{ssec:SoftwareInjections}

Different data sets have been prepared each containing one of
two different software-simulated CW signals without noise. 
A $\F$-statistic search has been conducted in each data set.
The software tools used for data production 
as well as for data analysis 
are part of the LSC Algorithm Library Applications~\cite{LALApps}.

The signal parameters defining the two simulated sources are given in
\Tref{t:swinj-signals}. In both cases the phase parameters
are chosen to be identical. These are also the same
phase parameters of the signal generating the hypersurfaces 
illustrated in \Fref{f:GCHypersurfaces}. 
Therefore, solely based on the global-correlation equations one expects
the $\F$-statistic to have a global large-value structure very similar to the one
shown by \Fref{f:GCHypersurfaces}. To investigate the impact of the antenna
pattern functions distinct amplitude parameters have been chosen here,
as given by \Tref{t:swinj-signals}.
For every set of simulated data the detector position refers 
to the LIGO Hanford 4-km (H1) detector, and the time of reference
is chosen to be global positioning system GPS time $793555944\,{\rm s}$ consistent
with \Fref{f:GCHypersurfaces}.

\begin{table}
\caption{\label{t:swinj-signals}
Amplitude and phase parameters introduced in \Sref{ssec:SimpleMatchedFilter} of 
the two software-simulated continuous gravitational-wave signals.}
\begin{ruledtabular}
\begin{tabular}{ccccccccc} 
Signal && $\;A_+\;$ && $\;A_\times\;$ && $\psi$ [rad] && $\Phi_0$ [rad] \\
\hline
1 && $1.0$ && $0.5$ && $1.0$ && $2.0$ \\
2 && $1.0$ && $0.0$ && $0.0$ && $0.0$\\
\hline\vspace{1pt}\\\hline
Signal && $\alpha_\sig$ [rad] && $\delta_\sig$ [rad] && $f_\sig$ [Hz] && $\fkdot{1}_\sig$ [Hz/s]\\
\hline
1 and 2 && $2.0$ && $-1.0$ && $100.0$ && $-10^{-10}$ \\
\end{tabular}
\end{ruledtabular}
\end{table}
\begin{table}
\caption{\label{t:swinj-data} Comparison of  predictions by the 
global-correlation equations based the simplified detection statistic~$\Xstat$ 
with the fully coherent $\F$-statistic search results using data sets containing 
software-simulated continuous gravitational-wave signals. 
The search labels, listed in the first column, containing the number 1 (number 2)
refer to data sets containing only Signal~1 (Signal~2), whereas labels with different letters
correspond to different observation times, as can be seen from the second column. 
From the obtained results the {\it maximum} relative 
deviations from the predicted frequencies are specified in the third column.}
\begin{ruledtabular}
\begin{tabular}{@{}llllcc}
Search&& Coherent  obser-&& Maximum deviation \\
label &&  vation time $T$  && $|f_{\F}-f_{\Xstat}|/f_{\Xstat}$\\
\hline
A1 && 10 hours && $2.5\times10^{-6}$\\ 
B1 && 1 sidereal day && $2.2\times10^{-6}$\\ 
C1 && 30 hours && $2.0\times10^{-6}$ \\ 
\vspace{5pt}
D1 && 2 sidereal days && $3.2\times10^{-6}$\\
A2 && 10 hours && $2.6\times10^{-6}$ \\ 
B2 && 1 sidereal day && $2.4\times10^{-6}$ \\ 
C2 && 30 hours && $2.3\times10^{-6}$\\ 
D2 && 2 sidereal days && $3.5\times10^{-6}$ \\ 
\end{tabular}
\end{ruledtabular}
\end{table}

As listed by \Tref{t:swinj-data}, for each data set containing the same
signal different searches have been conducted, where the coherent 
observation time~$T$ has been varied between from 10~hours up to 2~sidereal days.
In each search consisting of evaluating the $\F$-statistic on a grid of
templates, an isotropic sky grid with equatorial spacing of $0.02$~rad, 
spacings of~$1/(2T)$  in the frequency-interval~$f\in[99.8,100.2]\,\Hz$ 
and a fixed spin-down template of $\fkdot{1} = -10^{-11}\,\Hz/{\rm s}$ have 
been employed.

First we compare the results of the searches with the
prediction by the global-correlation~\Eref{e:C1}
forming the hypersurface~$\Hyps_1$, shown in \Fref{f:dsurfs1}. 
For each sky position and spin-down template-grid point where the $\F$-statistic search
reported a candidate event with frequency~$f_\F$,
one can calculate the relative deviation to the predicted frequency~$f_{\Xstat}$
obtained from \Eref{e:C1} by $| f_{\F} - f_{\Xstat} | / f_{\Xstat}$.
For each search the {\it maximum} relative deviations between 
predicted~$f_{\Xstat}$ and measured~$f_\F$ over the entire sky
(and spin-down) are specified in \Tref{t:swinj-data}, and are of order~$10^{-6}$.
The corresponding average relative deviations are typically even
one order of magnitude smaller. As in the simplified phase 
model~\eref{e:orbital-phase} the Earth's spinning component has been 
neglected, here the observed frequency-deviations are consistent with 
the fact that the relative corrections to the frequency modulation originating 
from the Earth's spinning velocity are of 
magnitude~${v}_{\rm spin}/c \approx10^{-6}$.

\begin{figure*}
	\center
	\subfigure[
	Results of $\F$-statistic search~A1 ($T=10\,{\rm h}$).
	\label{f:SoftSignalA1}]
	{\includegraphics[width=0.8\columnwidth]
	{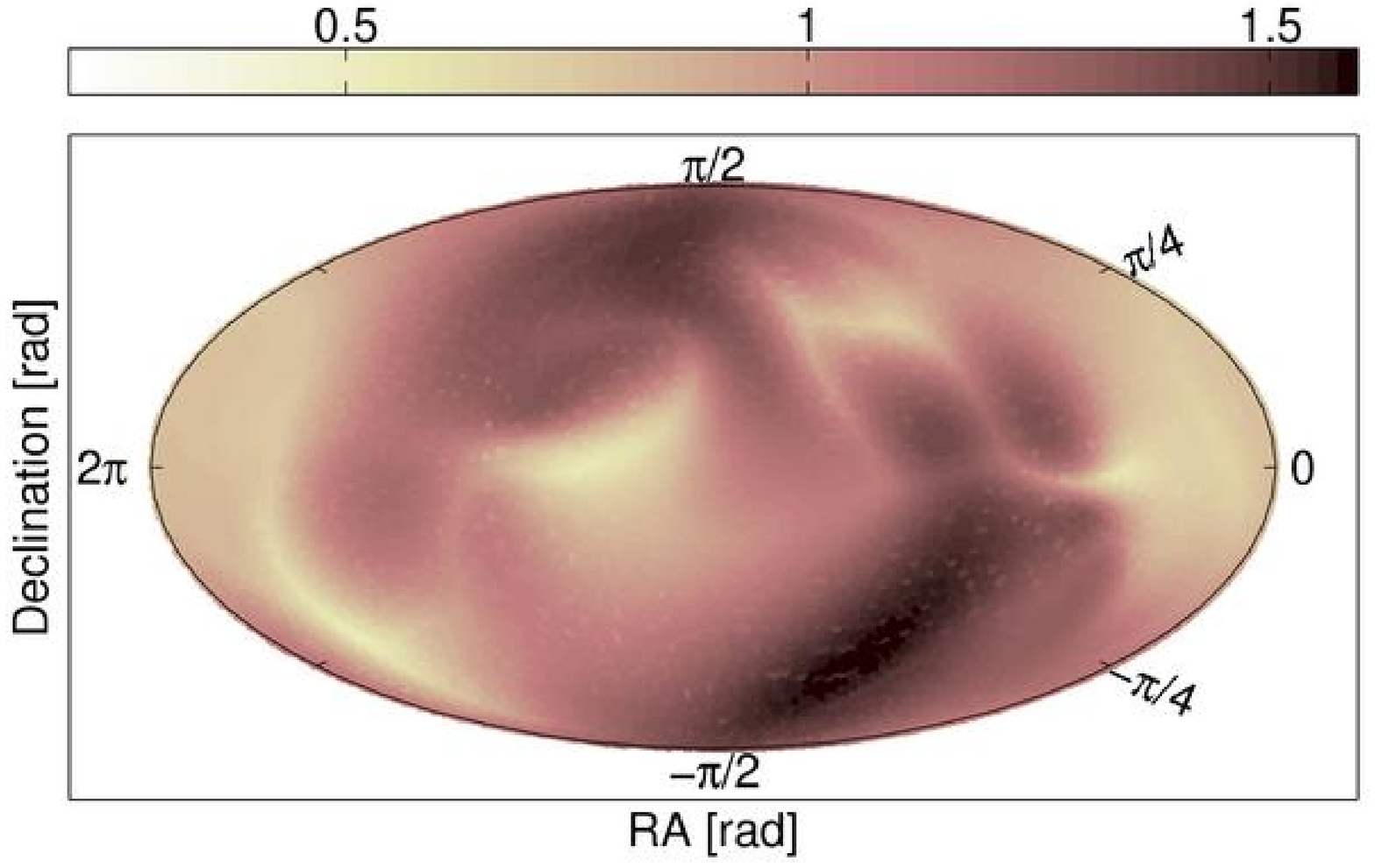}}\qquad\qquad
	\subfigure[
	Results of $\F$-statistic search~A2 ($T=10\,{\rm h}$).
	\label{f:SoftSignalA2}]
	{\includegraphics[width=0.8\columnwidth]
	{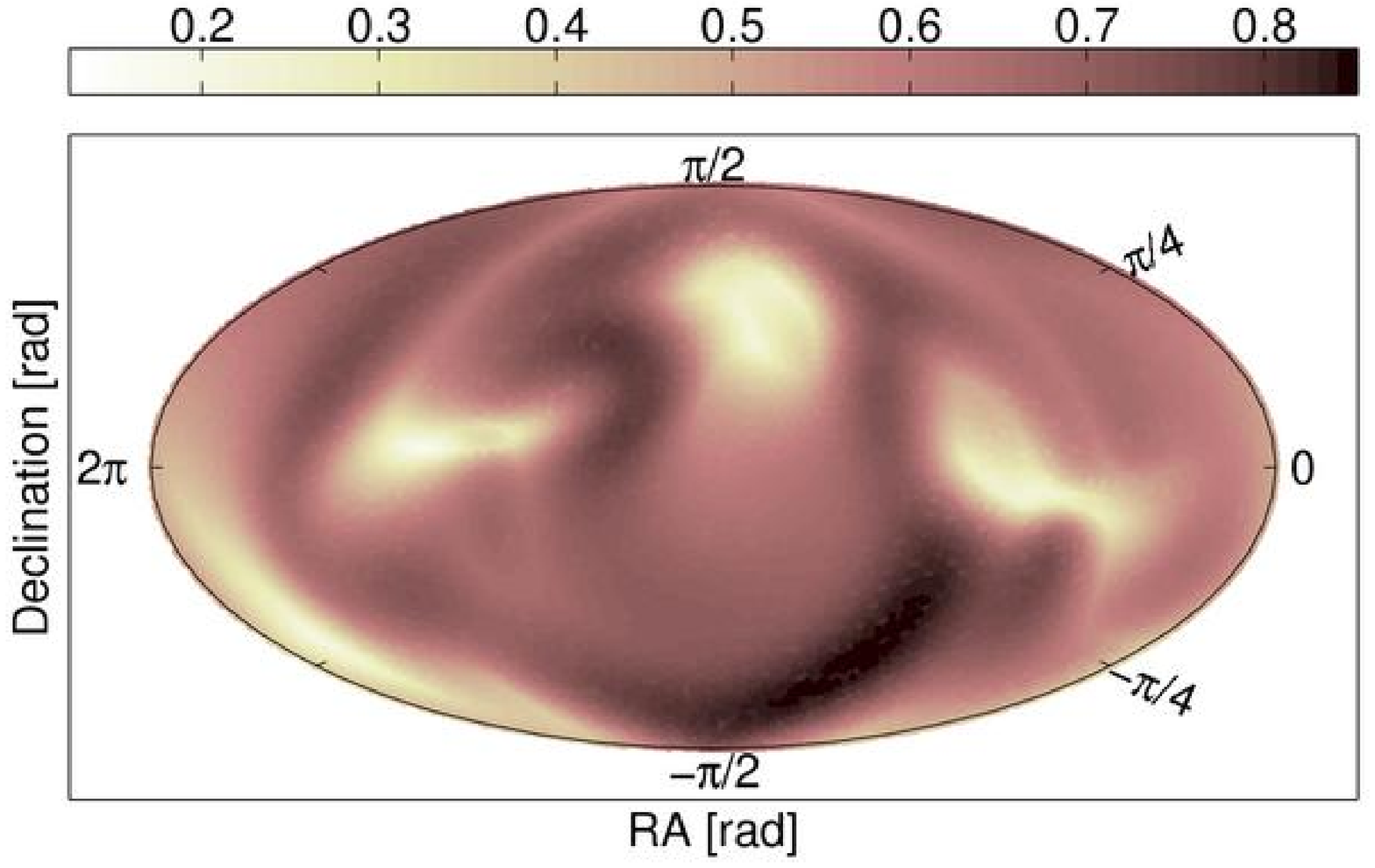}}\\
	\subfigure[
	Results of $\F$-statistic search~B1 ($T=1\,{\rm sd}$).
	\label{f:SoftSignalB1}]
	{\includegraphics[width=0.8\columnwidth]
	{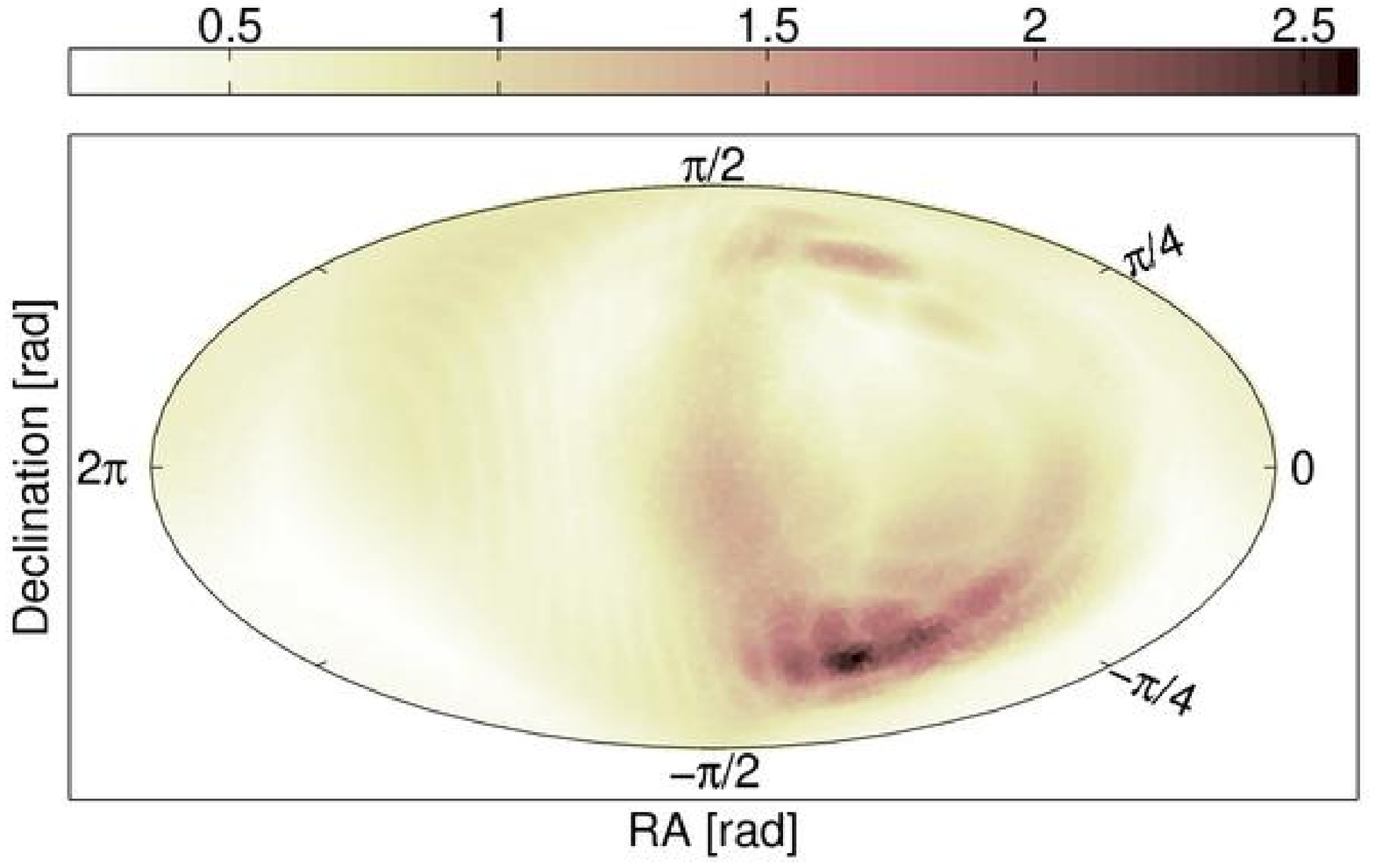}}\qquad\qquad
	\subfigure[
	Results of $\F$-statistic search~B2 ($T=1\,{\rm sd}$).
	\label{f:SoftSignalB2}]
	{\includegraphics[width=0.8\columnwidth]
	{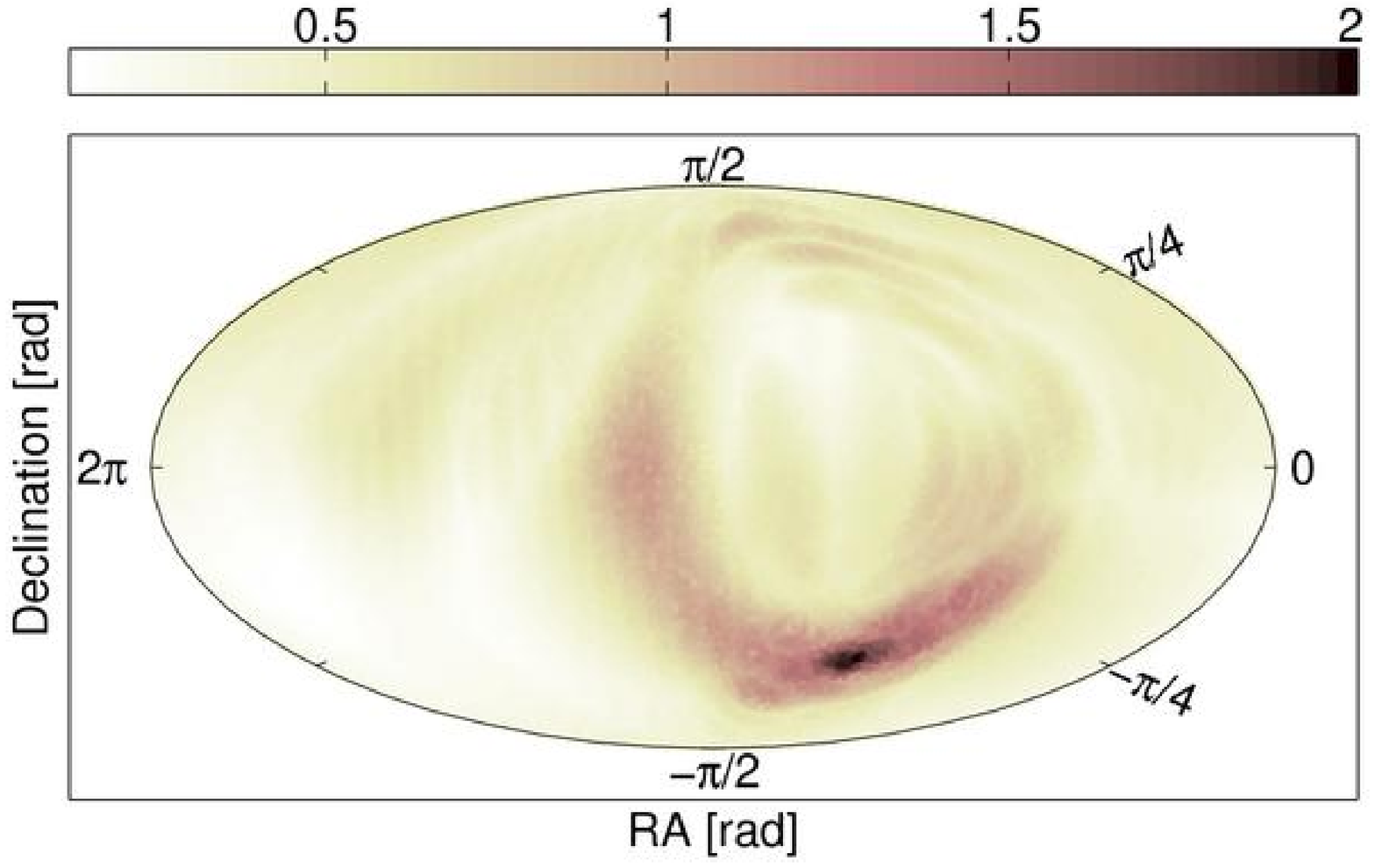}}\\
	\subfigure[
	Results of $\F$-statistic search~C1 ($T=30\,{\rm h}$).
	\label{f:SoftSignalC1}]
	{\includegraphics[width=0.8\columnwidth]
	{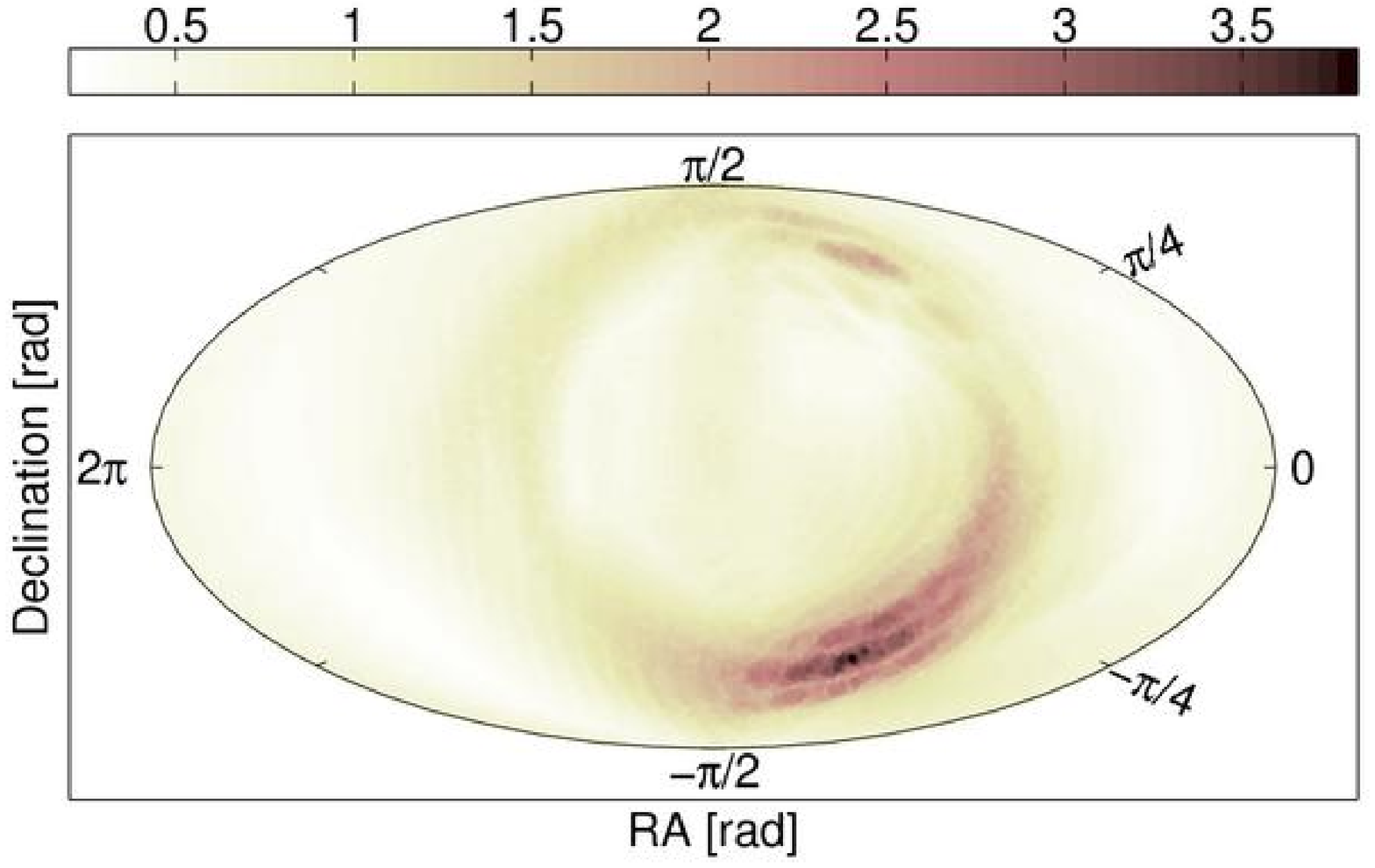}}\qquad\qquad
	\subfigure[
	Results of $\F$-statistic search~C2 ($T=30\,{\rm h}$).
	\label{f:SoftSignalC2}]
	{\includegraphics[width=0.8\columnwidth]
	{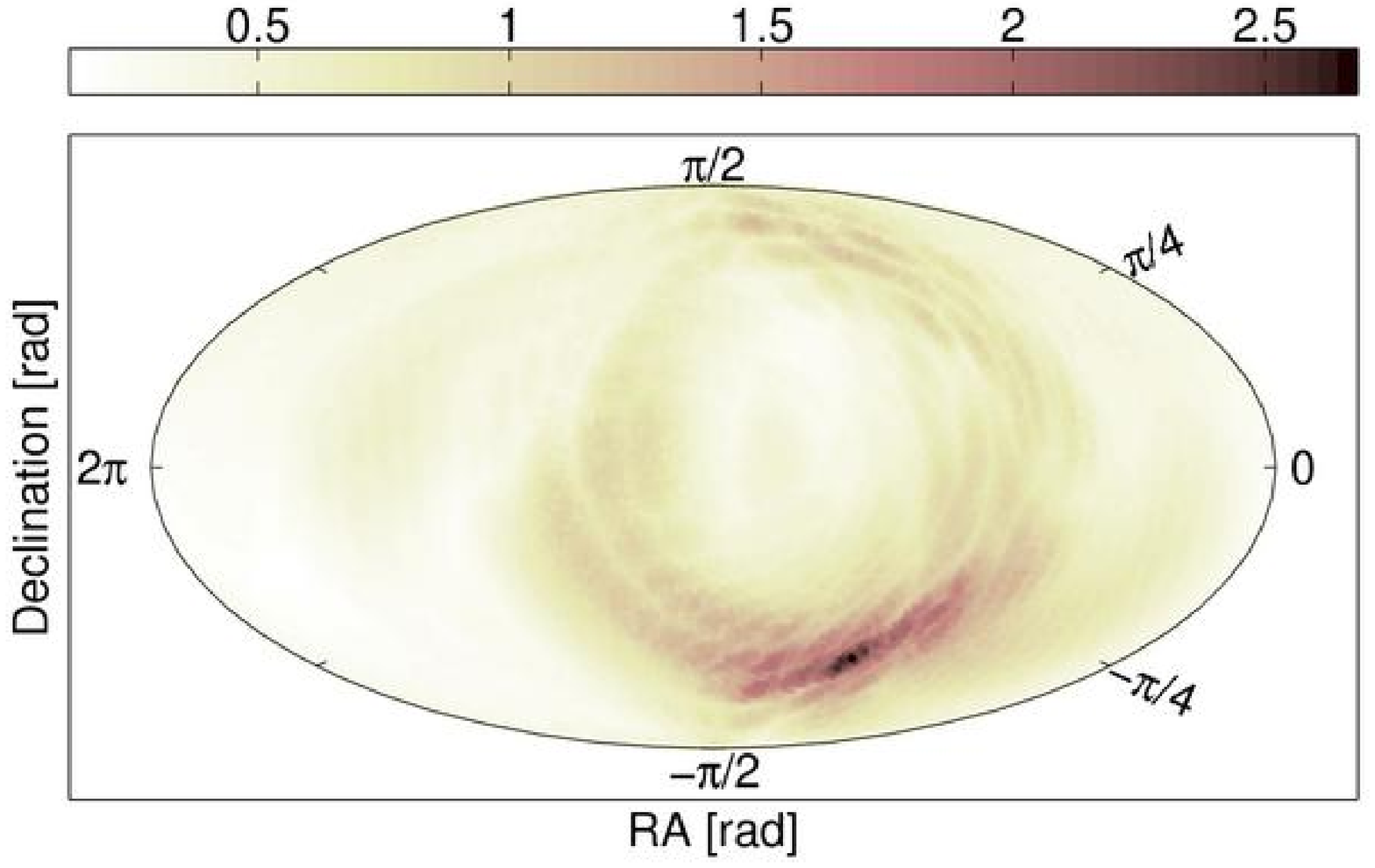}}\\
	\subfigure[
	Results of $\F$-statistic search~D1 ($T=2\,{\rm sd}$).
	\label{f:SoftSignalD1}]
	{\includegraphics[width=0.8\columnwidth]
	{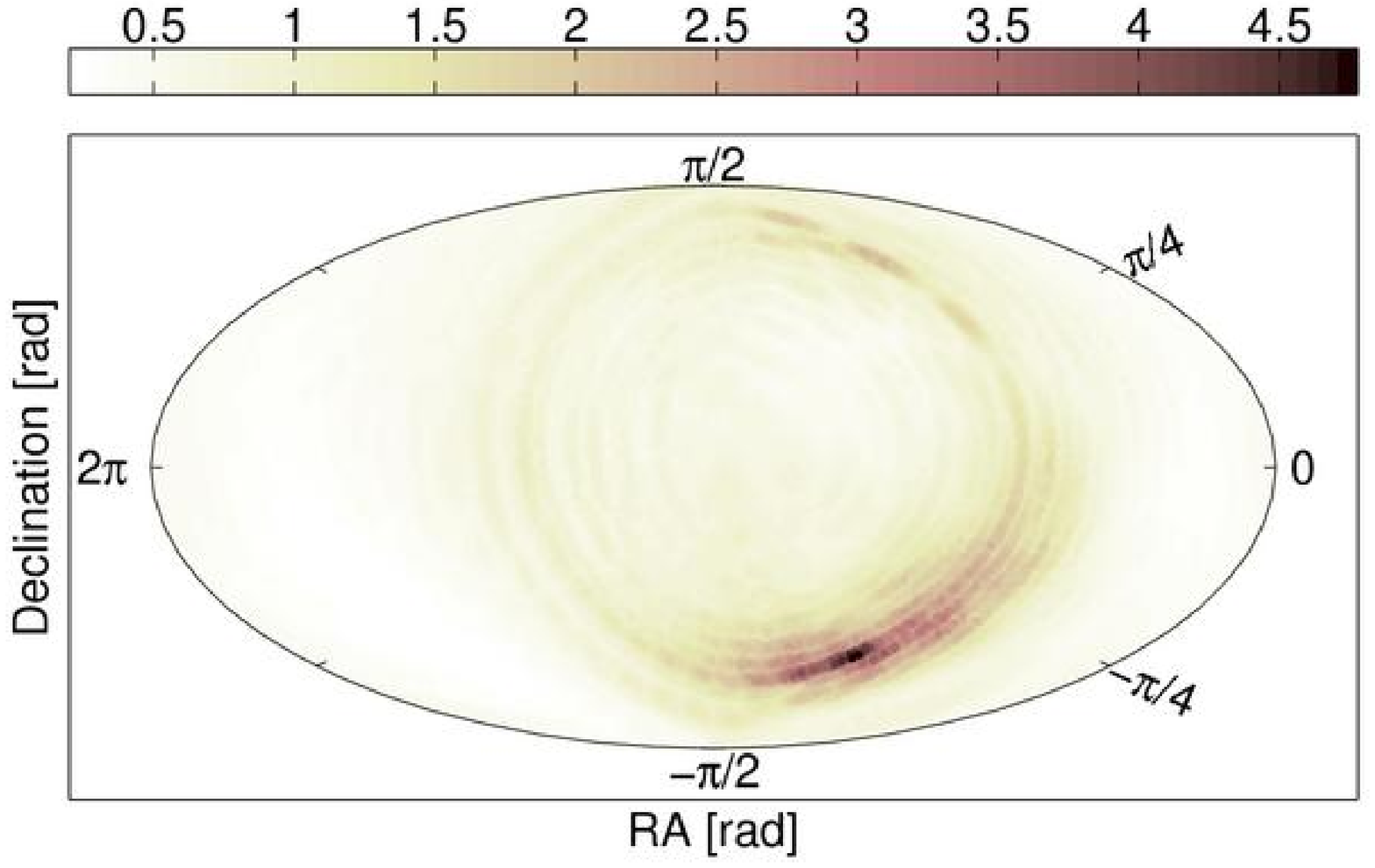}}\qquad\qquad
	\subfigure[
	Results of $\F$-statistic search~D2 ($T=2\,{\rm sd}$).
	\label{f:SoftSignalD2}]
	{\includegraphics[width=0.8\columnwidth]
	{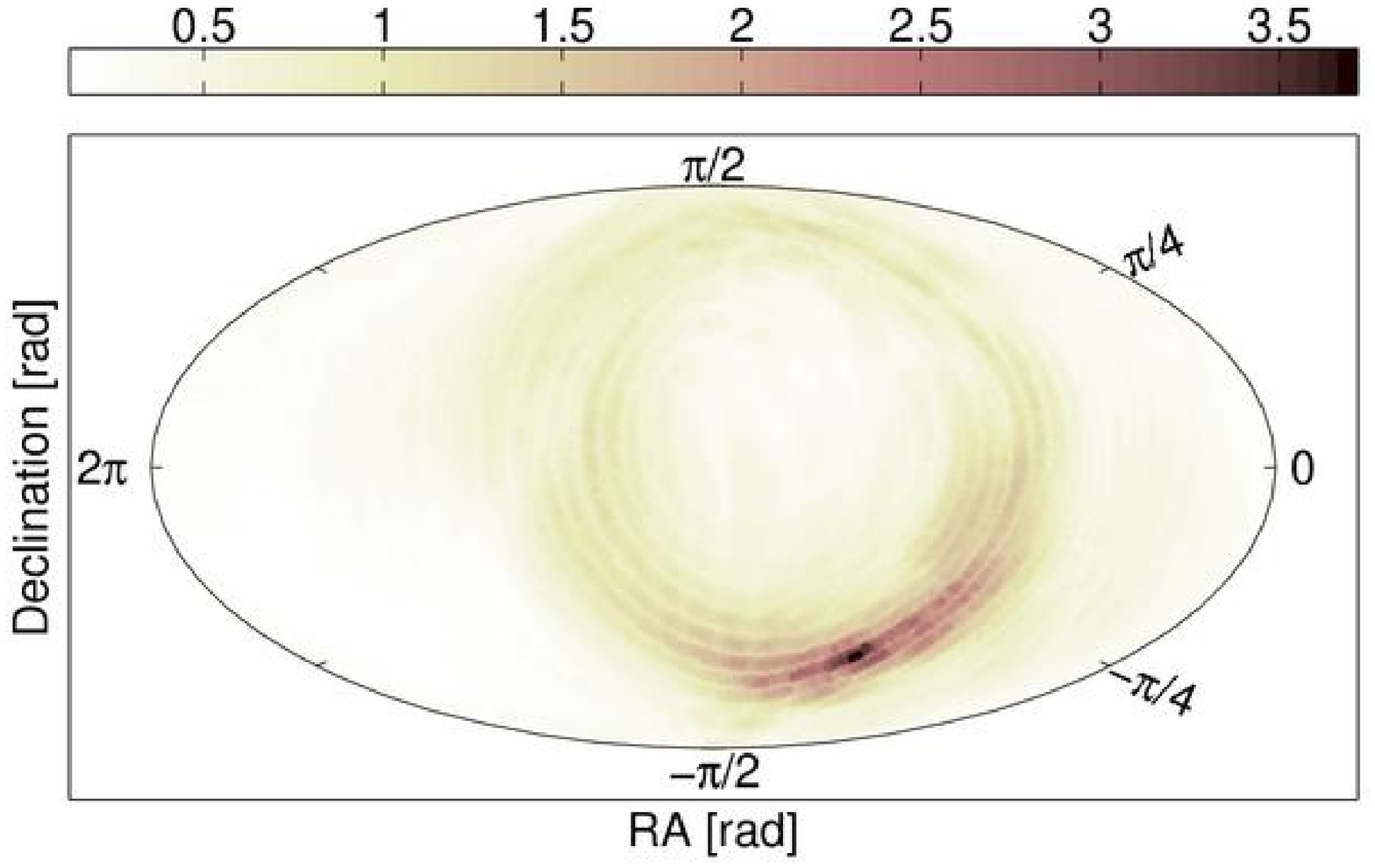}}
	\caption{
	Hammer-Aitoff sky projections of results from fully 
	coherent $\F$-statistic searches in the data sets 
	described by \Tref{t:swinj-data}. 
	Each data set contained one of the two software-simulated CW 
	sources defined in \Tref{t:swinj-signals}, where
	both signals~1 and~2 have identical phase 
	parameters, but different amplitude parameters.
	The plots of the left (right) column show candidate events registered 
	in the different searches A1,B1,C1 and D1 (A2,B2,C2 and D2) above 
	a threshold of $2\F \ge 2000$ ($2\F \ge 1250$). The colorbar indicates the 
	values of $2\F \times10^{-4}$.
	The corresponding analytical prediction by the global-correlation equations 
         has been illustrated earlier by \Fref{f:dsurfs5}.
	\label{f:SoftwInj}}
\end{figure*}

Finally, \Fref{f:SoftwInj} presents the results of each $\F$-statistic search
projected on the sky.  As we are only interested in the loudest 
$\F$-statistic values, only strong candidate events from the ones 
reported in the different searches A1,B1,C1 and D1 above 
a threshold of $2\F \ge 2000$ are shown, and
for the searches A2, B2, C2 and D2 a threshold
of $2\F \ge 1250$ is set. These thresholds were chosen to reduce
the data volume to process and are based the
expected global maximum values of $2\F$. 
These thresholds guarantee that  candidate events with
$\F$-statistic values at least larger than $\approx 15\%$ of
the global maximum are retained in each search. This thresholding
is justified because comparing candidate events with very small detection-statistic 
values to the global maximum are not of interest in this work.

As shown earlier in \Fref{f:GCHypersurfaces}, in the 
two-dimensional sky projection the intersection curve of 
hypersurfaces~$\Hyps_1$ and~$\Hyps_2$ approximately coincides
with the contours (of constant $f$ and $\fkdot{1}$) of~$\Hyps_2$ in the sky.
Thus, the dark circle in \Fref{f:dsurfs5} representing the 
contours of hypersurface~$\Hyps_2$ also describes the predicted global
maximum structure of the detection statistic in the sky. In fact,
this prediction is actually observed in qualitatively good agreement with
the  $\F$-statistic search results shown in \Fref{f:SoftwInj}.
One finds that for coherent observation times less than one sidereal day 
(in searches A1 and A2), the locations of the predicted global maximum structure 
are only faintly visible in the results, because this feature is still hidden due to 
the Earth's spin component [see Figures~\ref{f:SoftSignalA1}~and~\subref{f:SoftSignalA2}].
For coherent observation times beyond one period of the Earth's 
spinning motion, the results clearly show the locations of large $\F$-values
as predicted [see Figures~\ref{f:SoftSignalB1}
-- \subref{f:SoftSignalD2}]. 
As will be discussed later in \Sref{sec:Xspin},
the Earth's spinning motion only varies the detection statistic 
within the global-correlation-equations 
predicted (and observed) regions.

\subsection{Comparison with the $\F$-statistic for a hardware-injected signal
in detector noise}
\label{ssec:HardwareInjection}

For so-called ``hardware injections'' simulated signals are physically 
added into the detector control systems to produce instrumental signals that
are similar to those that are expected to be produced by astrophysical
sources of gravitational waves. Through magnetic coil actuators
the interferometer mirrors are made to physically move as if a
gravitational wave was present.  

Here, we choose $30\,{\rm hours}$ of data (lying within a time span of 
less than $38\,{\rm hours}$) containing a hardware injection 
from LIGO's fourth science run (S4) of the LIGO Livingston 4-km (L1) detector.
The GPS time of reference is $795408715\,{\rm s}$. During this data segment the
hardware injection was activated $99.4\%$ of the time.
The signal's phase parameters are defined by $\alpha_\sig=3.7579\,{\rm rad}$, 
$\delta_\sig=0.0601\,{\rm rad}$, $f_\sig = 575.163636\,\Hz$, 
and $\fkdot{1}_\sig = -1.37\times 10^{-13} \Hz / {\rm s}$, at the segment's starting time.
The amplitude parameters are $A_+=7.48\times 10^{-24}$, 
$A_\times  = -7.46\times 10^{-24}$,
$\psi=-0.22\,{\rm rad}$ and the initial phase is $\Phi_0=4.03\,{\rm rad}$. 
In this segment of data an all-sky $\F$-statistic search has been carried out 
in the frequency interval of $f\in[575.048,575.221]\,\Hz$, and in the
spin-down range $\fkdot{1} \in [-1.04\times 10^{-9}, 1.04\times10^{-10}]\,\Hz/{\rm s}$.
The grid of templates employed spacings of~$5.70\times10^{-6}\,\Hz$ in 
$f$-direction, separations of $3.23\times10^{-10}\,\Hz/{\rm s}$ in $\fkdot{1}$-direction,
and an isotropic sky grid with equatorial spacing of $0.03\,{\rm rad}$.

\begin{figure}
	\subfigure[Results of a coherent all-sky $\F$-statistic search
	 in a data set containing a hardware-injected signal.
	The colorbar indicates the values of $2\F$.
	\label{f:hwinj1}]
	{\includegraphics[width=0.9\columnwidth]
	{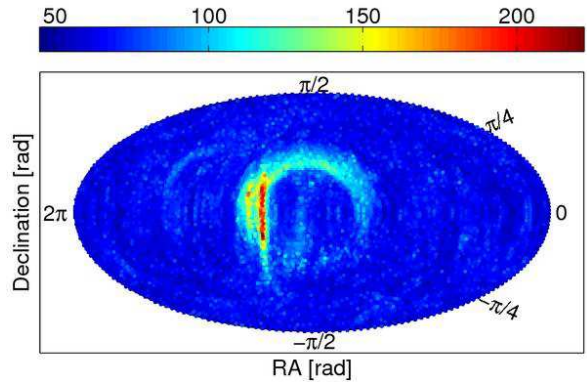}}
	\\
	\subfigure[Prediction of the global maximum structure of the 
	detection-statistic based on the global-correlation hypersurfaces.
	\label{f:hwinj2}]
	{\includegraphics[width=0.9\columnwidth]
	{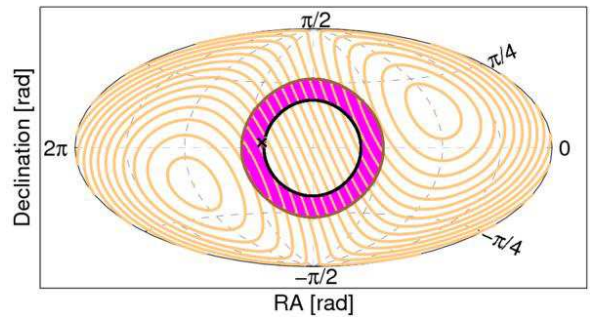}}
	\caption{Comparing the  $\F$-statistic results~\subref{f:hwinj1} 
	for a hardware-injected CW signal with the theoretical 
	prediction~\subref{f:hwinj2} by the global-correlation equations.
	Both plots show Hammer-Aitoff projections of the sky.
	The sky location of the simulated signal is represented by
	the black cross in~\subref{f:hwinj2}.
	The magenta region in~\subref{f:hwinj2} represents the
	predicted structure of maximum detection statistic~$\Xstat$ and
	is observed to agree qualitatively well with the actual $\F$-statistic 
	results~\subref{f:hwinj1} from the hardware injection in real detector data.
	\label{f:HardwareInjection}} 
\end{figure}

In \Fref{f:HardwareInjection} the results of the actual $\F$-statistic search 
are shown and compared to the prediction by the global-correlation
hypersurfaces. As the search covers a range of spin-down templates
there will distinct contours of  hypersurface~$\Hyps_2$ corresponding to each
$\fkdot{1}$-template. Therefore, the structure of maximum detection statistic~$\Xstat$
is expected to be an annulus in the sky referring to the magenta region
in~\Fref{f:hwinj2}. This annulus is framed by the different contour lines of~$\Hyps_2$
which correspond to the minimum (dark-brown colored) 
and maximum (black colored) value of~$\fkdot{1}$ searched.
Qualitatively, the large-value structure of~$\F$ that is visible, is in good agreement with 
the predicted structure based on the global-correlation hypersurfaces.
In~\Fref{f:hwinj1} the $\F$-statistic results only faintly reproduce
the entire predicted structure because of the Earth's spinning motion.
\Sref{sec:Xspin} will discuss and explain analytically how the 
Earth's spinning motion varies the detection statistic within the
regions determined by the global-correlation hypersurfaces.

\section{Vetoing instrumental noise artifacts}
\label{sec:veto}

A typical feature of the data from an interferometric gravitational-wave 
detector are narrow-band noise artifacts, so-called ``lines",
which are of instrumental origin. As a consequence, the results of a search for
continuous gravitational-wave sources contain instrumental
artifacts that in some respects mimic CW signals.  But these artifacts
tend to cluster in certain regions of parameter space. 
For the case of {\it incoherent} searches as reported in~\cite{pshS4:2008}, 
candidate events in such parameter-space regions were identified and
discriminated. Here, we propose a similar veto method also suitable
for {\it coherent} CW searches, such as those found in~\cite{eahS4:2008}.
Using the global-correlation equations found in this work we aim 
to identify those regions in parameter space where instrumental noise 
lines can imitate a real signal by producing large detection-statistic values. 
In such a case these candidate events could automatically be vetoed.  

For simplicity we consider the same four-dimensional parameter space 
as used in Section~\ref{ssec:geometry} consisting of $\{f,\fkdot{1}, \vec{n}\}$
and that $T$ has a value, such that third-order contributions to $|\Xamp|$ 
are insignificant (cf. \Sref{ssec:third-order-est}).
Thus, the global-correlation equations of relevance are the same as in
the example studied in \Sref{ssec:example}.
These were  given by Equations~(\ref{e:C1}) and (\ref{e:C2}). 
It is obvious that the frequency of a stationary instrumental line
is independent of the Earth's position in its orbit around
the sun. This decoupling is achieved by setting $\vec{n}_\sig = \vec 0$. 
For a stationary instrumental line originating from the detector 
it also holds that $\fkdot{1}_\sig = 0$. In this case the 
constants~$K_{1,2}$ in Equations~(\ref{e:C1}) and~(\ref{e:C2}) 
simplify to $K_1=f_\sig$ and $K_2=0$. 
Thus, the two relevant global-correlation equations are of the 
following form,
\begin{subequations}\label{e:N12}
\begin{eqnarray}
  f + f \, \xil{1} \cdot \vec n 
  + \fkdot{1}  \,\vec{\xi} \cdot \vec n &=& f_\sig ,
  \label{e:N1}\\
  \fkdot{1} + f \, \xil{2} \cdot \vec n 
  + 2 \fkdot{1}  \,\xil{1} \cdot \vec n &=&  0 
  \label{e:N2} .
\end{eqnarray}
\end{subequations}
As stated earlier, for a given $\fkdot{1}$, \Eref{e:N1} describes a three-dimensional 
hypersurface in the subspace $\{f,\vec n\}$. On this hypersurface
the detection statistic will attain its maximum along the intersection curve with
the hypersurface described by \Eref{e:N2}. In a projection into the 
sky subspace this intersection curve approximately coincides with the
contours (of constant~$f$ and~$\fkdot{1}$) of hypersurface~\eref{e:N2}.
Therefore \Eref{e:N2} describes the region in the sky 
for which potential CW signals do not produce a modulation pattern 
that would distinguish them from an instrumental line.

Using this knowledge one can discriminate (veto) candidate events
which satisfy \Eref{e:N2}. As the resolution in parameter space 
is finite, we formulate the following {\it veto~condition}:
\begin{equation}
  \left|  \,  \fkdot{1} + f \, \xil{2} \cdot \vec n 
  + 2 \fkdot{1}  \,\xil{1} \cdot \vec n\, \right| < \varepsilon ,
  \label{e:veto-cond}
\end{equation}
the tolerance-parameter $\varepsilon > 0$ can be understood as
\begin{equation}
  \varepsilon = \frac{\Delta f_{\rm cell}}{\Delta T}\,N_{\rm cell} ,
\end{equation}
where $\Delta f_{\rm cell}$ denotes the resolution in the frequency-direction
(width of cells), $N_{\rm cell}$ the number of cells one tolerates during
a characteristic length of time $\Delta T$.

One can visualize and calculate the volume of the region in
four-dimensional parameter space which is excluded by this veto.  For
a given source sky position, \Eref{e:veto-cond} is linear in 
$f$ and $\fkdot{1}$. Thus, for fixed sky position
${\vec{n}}$, the veto condition defines two parallel lines in the 
$\{f,\fkdot{1}\}-{\rm plane}$. Candidate events which lie
in the region between the lines are discarded (vetoed).  Candidate events
which lie outside this region are retained (not vetoed).  The
locations of these two lines in the $\{f,\fkdot{1}\}-{\rm plane}$ depends upon
the sky position.  The fractional volume excluded by the veto depends
upon whether or not (as the source position varies over the sky) the
excluded region between the lines lies inside or outside of the
boundaries of the search, or intersects it.  
Alternatively, for a given value of $f$ and $\fkdot{1}$, one can
calculate the portion of the sky which is excluded by the veto,
depending upon the ranges of parameter space searched.
The details of such a calculation for a particular search can be
found in Appendix A of~\cite{eahS4:2008}.

\Fref{f:NoiseLine} illustrates the veto method for an example noise line. Thereby, 
\Fref{f:NoiseLine1} presents the results of a fully coherent matched-filtering
search using the $\F$-statistic for a 30-hour observation time. The data set analyzed contains
a detector-noise line, which is in this case a violin mode resonance 
of the mode cleaner mirrors of the LIGO Hanford 4-km (H1) detector.
In \Fref{f:NoiseLine2}, a comparison with
the theoretical prediction by the global-correlation 
equations given by~\eref{e:N1} and~\eref{e:N2} is made
featuring a very good  agreement. Thus the veto will
be very efficient in excluding the parameter-space regions of
largest $2\F$-values produced by instrumental lines.

\begin{figure}
	\subfigure[Results of a coherent all-sky $\F$-statistic search
	in a data set containing an instrumental noise line.
	The colorbar indicates the values of $2\F$.
	\label{f:NoiseLine1}]
	{\includegraphics[width=0.9\columnwidth]
	{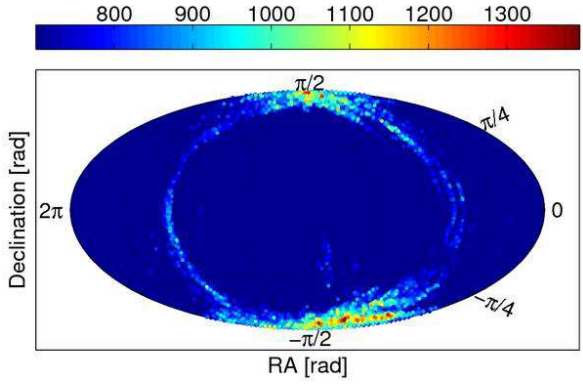}}
	\\
	\subfigure[Prediction of the global maximum structure of
	the detection statistic based on the global-correlation hypersurfaces.
	\label{f:NoiseLine2}]
	{\includegraphics[width=0.9\columnwidth]
	{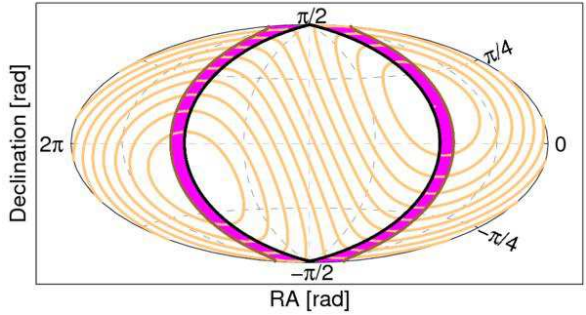}} 
	\caption{Comparing the results of a fully coherent matched-filtering
	search using the $\F$-statistic \subref{f:NoiseLine1} with the theoretical 
	prediction  \subref{f:NoiseLine2} by the global-correlation equations
	for a given instrumental-noise feature 
	mimicking a real CW signal. Both plots show Hammer-Aitoff projections of the sky.
	The all-sky search was carried out in a $0.5\,\Hz$ frequency-band 
	$f \in [568.0,568.5]\,\Hz$ and for a range of frequency time-derivatives
	 $\fkdot{1} \in [-3.63\times10^{-9},3.63\times10^{-10}]\,\Hz/{\rm s}$,  for an observation time
	 of $T=30\,{\rm h}$. The GPS time of reference is $795149016\,{\rm s}$.
	The upper plot~\subref{f:NoiseLine1} shows all candidate events reported by the search 
	above a detection-statistic threshold of approximately 50\% of the largest $2\F$-value
	found. The frequency of the instrumental-noise line present in this data set
	is a resonance violin mode of the mode cleaner mirrors of the 
	LIGO Hanford 4-km (H1) detector. 
	The magenta region in~\subref{f:NoiseLine2} of maximum expected
	detection statistic agrees well with the $\F$-statistic results~\subref{f:NoiseLine1}
	from the real instrumental line.
	\label{f:NoiseLine}} 
\end{figure}

Note that this veto excludes only the loudest 
candidate events  (of largest 2$\F$), but as 
shown in \Fref{f:NoiseLine} an instrumental
line is capable of contaminating large parts of the sky due to the
global correlations (depending on the search parameters).
Therefore in some cases it might be necessary to increase the
tolerance-parameter~$\varepsilon$ artificially to account for
these effects if necessary. However, in a $\F$-statistic search
one is interested in the strongest candidate events arising
from the background level. Thus in eliminating such
false instrumental-noise events the veto condition presented
is very efficient as these regions are well described. 
This veto method as presented here is applied in the Einstein@Home 
search described in~\cite{eahS4:2008}.

\section{Effects of the diurnal spinning motion of the Earth}
\label{sec:Xspin}
By considering the Earth's spinning motion in the phase model, we here investigate the variation of the 
detection statistic along the predicted global maximum structure by the global-correlation hypersurfaces.
In other words, given the solution $\Delta u_m=0$, which are the 
the global-correlation hypersurfaces, we study how the spin 
component $\phi_{\rm spin}(t)$
of \Eref{e:phase-spin-part} in the phase model~\eref{e:phase model} modulates the
detection statistic in the locations consistent with this solution.
In order to simplify this discussion, in what follows
frequency time-derivatives in the spin component $\phi_{\rm spin}(t)$ are ignored to
obtain
\begin{equation}
 \phi_{\rm spin}(t) \approx 2\pi\,f\,\frac{\vec r_{\rm spin} (t) \cdot \vec{n}}{c}.
 \label{e:phase-spin-f}
\end{equation}
Thus, the phase difference between the spin component of the signal~$\phi^\sig_{\rm spin}(t)$
and a template~$\phi_{\rm spin}(t)$ is given by
\begin{eqnarray}
  \Delta \phi_{\rm spin}(t) &\equiv& \phi^\sig_{\rm spin}(t) - \phi_{\rm spin}(t) \nonumber\\
  &=& 2\pi\,\frac{\vec r_{\rm spin}(t)}{c} \cdot \left( \vec{n}_\sig f_\sig - \vec{n} f \right)\nonumber\\
  &=& \frac{\vec r_{\rm spin}(t)}{c} \cdot \Delta\vec{k},
  \label{e:D-phi-spin}
\end{eqnarray}
where we defined the vector 
\begin{equation}
  \Delta\vec{k} \equiv 2\pi\,\left(\vec{n}_\sig f_\sig - \vec{n} f \right).
  \label{e:Dk}
\end{equation}
Taking into account $\Delta \phi_{\rm spin}(t)$ in the detection-statistic amplitude 
and provided that $\Delta u_m=0$, one has to compute the 
following integral
\begin{equation}
  X_{\rm spin} \equiv \frac{1}{T} \int_{-T/2}^{T/2} \, \rme^{i \Delta\phi_{\rm spin}(t)} \, \rmd t \,.
  \label{e:Xspin}
\end{equation}

For observation times $T$ relevant to CW searches (of order days), the phase modulation due to the 
spinning motion of the Earth is oscillatory, because it has a period of one sidereal day
$\Omega_{\rm spin} = 2\pi/1\,\rm sd$.
Therefore, in order to evaluate~\eref{e:Xspin}
we follow a route previously taken in~\cite{jotania:1996,prixitoh:2005}, which
makes use of the Jacobi--Anger identity:
\begin{equation}
  \rme^{iz\cos \theta} = \sum_{n=-\infty}^{\infty}i^n J_n(z)\,\rme^{in\theta},
\end{equation}
where $J_n(z)$ is the $n$-th Bessel function of the first kind. 
This identity allows to expand exponentials of trigonometric functions in the basis of their harmonics.
To employ the Jacobi--Anger expansion we rewrite \Eref{e:D-phi-spin} by approximating 
the diurnal detector motion due to the Earth's rotation to be circular,
\begin{eqnarray}
  \Delta \phi_{\rm spin}(t) &=& \frac{R_{\rm E}}{c}\biggl[{\Delta k}_{\|}\sin\lambda\nonumber\\
  & & + {\Delta k}_{\perp}\,\cos\lambda\,\cos(\phi_0+\Omega_{\rm spin}t)\biggr],
\end{eqnarray}
where $R_{\rm E}$ is the radius of the Earth, $\lambda$ is the latitude of the detector, 
${\Delta k}_{\|}$ is the absolute value of the component of the vector $\Delta\vec{k}$ parallel 
to the rotation axis, ${\Delta k}_{\perp}$ is the absolute value of the component of $\Delta\vec{k}$ 
orthogonal to the rotation axis, and $\phi_0$ is determined by $\Delta\vec{k}$ at $t=0$.
Defining 
\begin{eqnarray}
  \Delta\phi_{{\rm spin},\|} &\equiv& \frac{R_{\rm E}}{c}\, {\Delta k}_{\|}\sin\lambda,\\
  \Delta u_{\rm spin} &\equiv& \frac{R_{\rm E}}{c}\, {\Delta k}_{\perp}\cos\lambda
  \label{e:uspin},
\end{eqnarray}
we apply the Jacobi--Anger identity:
\begin{equation}
  \rme^{i \Delta\phi_{\rm spin}(t)}  = \rme^{i \Delta\phi_{{\rm spin},\|}} 
  \sum_{n=-\infty}^{\infty}i^n\,\rme^{in\phi_0}J_n( \Delta u_{\rm spin})\rme^{in\Omega_{\rm spin}t}.
\end{equation}
Substituting this expression into \Eref{e:Xspin} and taking the modulus,
one obtains
\begin{equation}
  |X_{\rm spin}| = \left| \sum_{n=-\infty}^{\infty}i^n\,\rme^{in\phi_0}J_n( \Delta u_{\rm spin})\,
  \sinc\left(\frac{n\Omega_{\rm spin}T}{2}\right)\right|.
  \label{e:mod-Xspin}
\end{equation}

\begin{figure}
	\subfigure[LIGO Hanford detector]{
	\includegraphics[width=0.9\columnwidth]
	{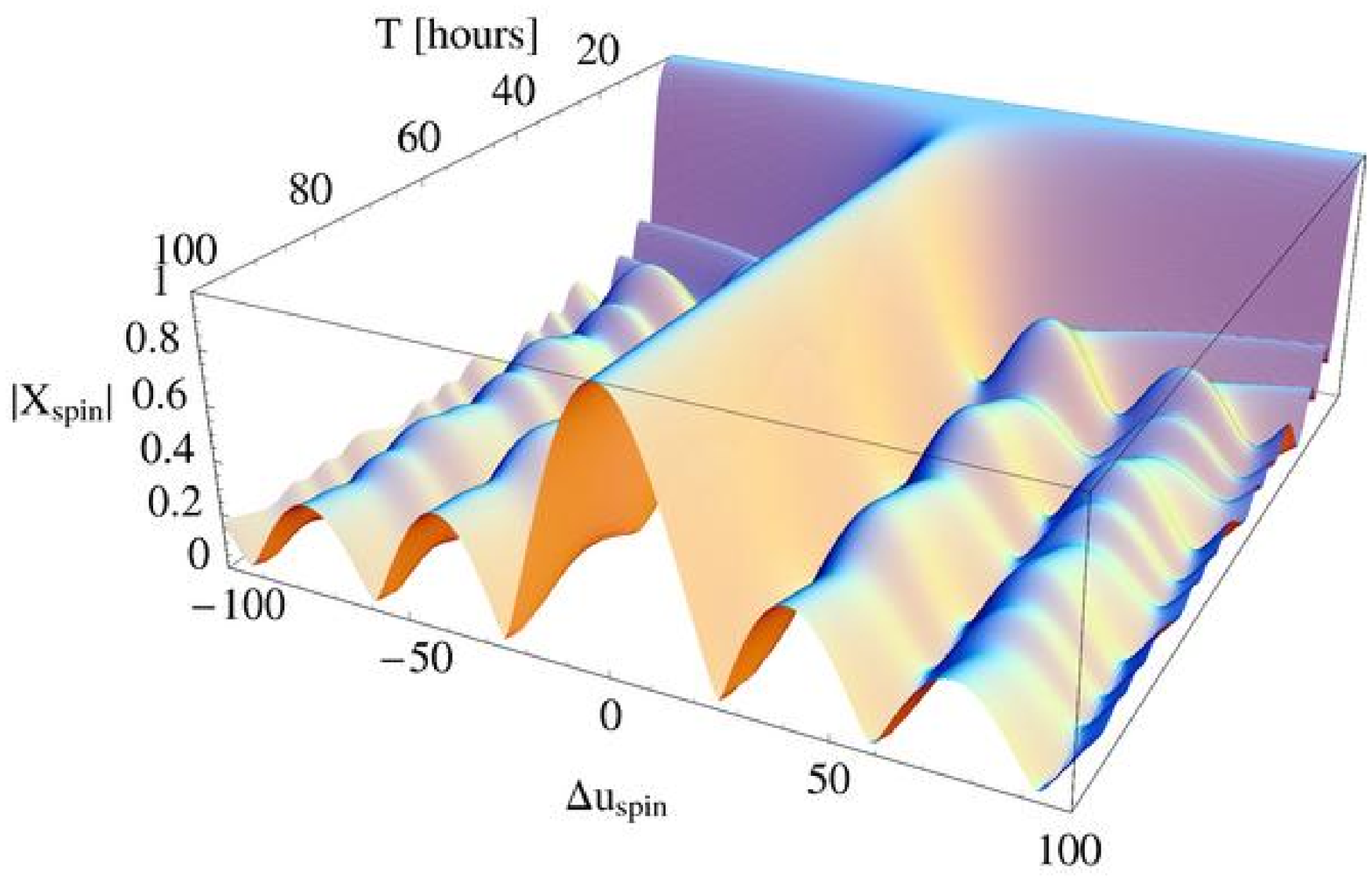}}\qquad
	\subfigure[LIGO Livingston detector]{
	\includegraphics[width=0.9\columnwidth]
	{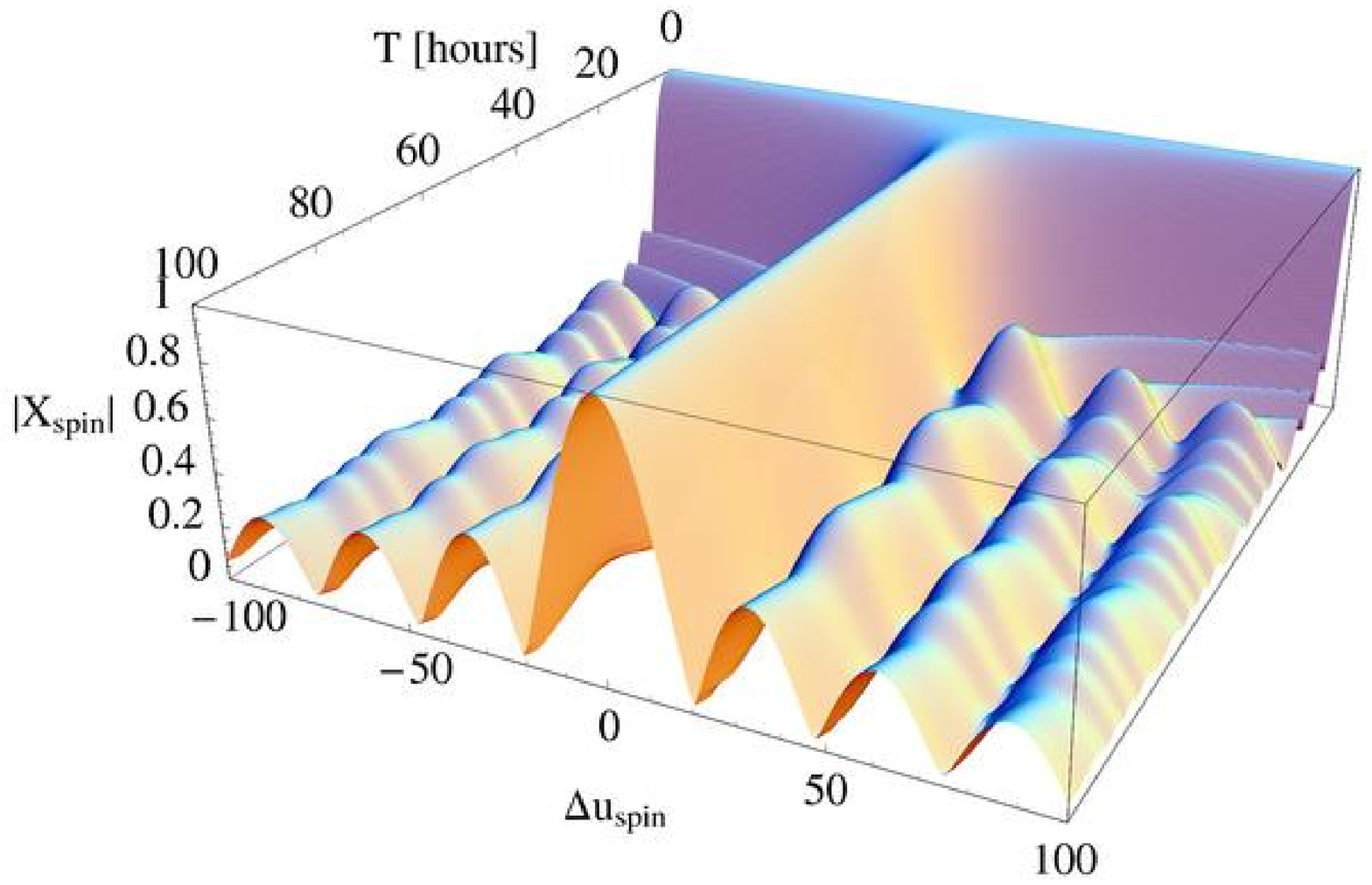}}
	\caption{(Color online) The simplified detection-statistic amplitude $|X_{\rm spin}|$ for
	phase-mismatch only in the spin component of the phase model as a function
	of observation time $T$ and dimensionless parameter-mismatch $\Delta u_{\rm spin}$ as defined
	in \Eref{e:uspin}, for the two LIGO detectors. For observation times beyond 
	$T\gtrsim2\pi/\Omega_{\rm spin}$, a good approximation of $|X_{\rm spin}|$ is given
	by the dominant term $|J_0( \Delta u_{\rm spin})|$.
	\label{f:Xspin}}
\end{figure}

\begin{figure*}
	\subfigure[The software-injected signal.\label{f:Xspin-swinj}]{
	\includegraphics[width=0.6\columnwidth]
	{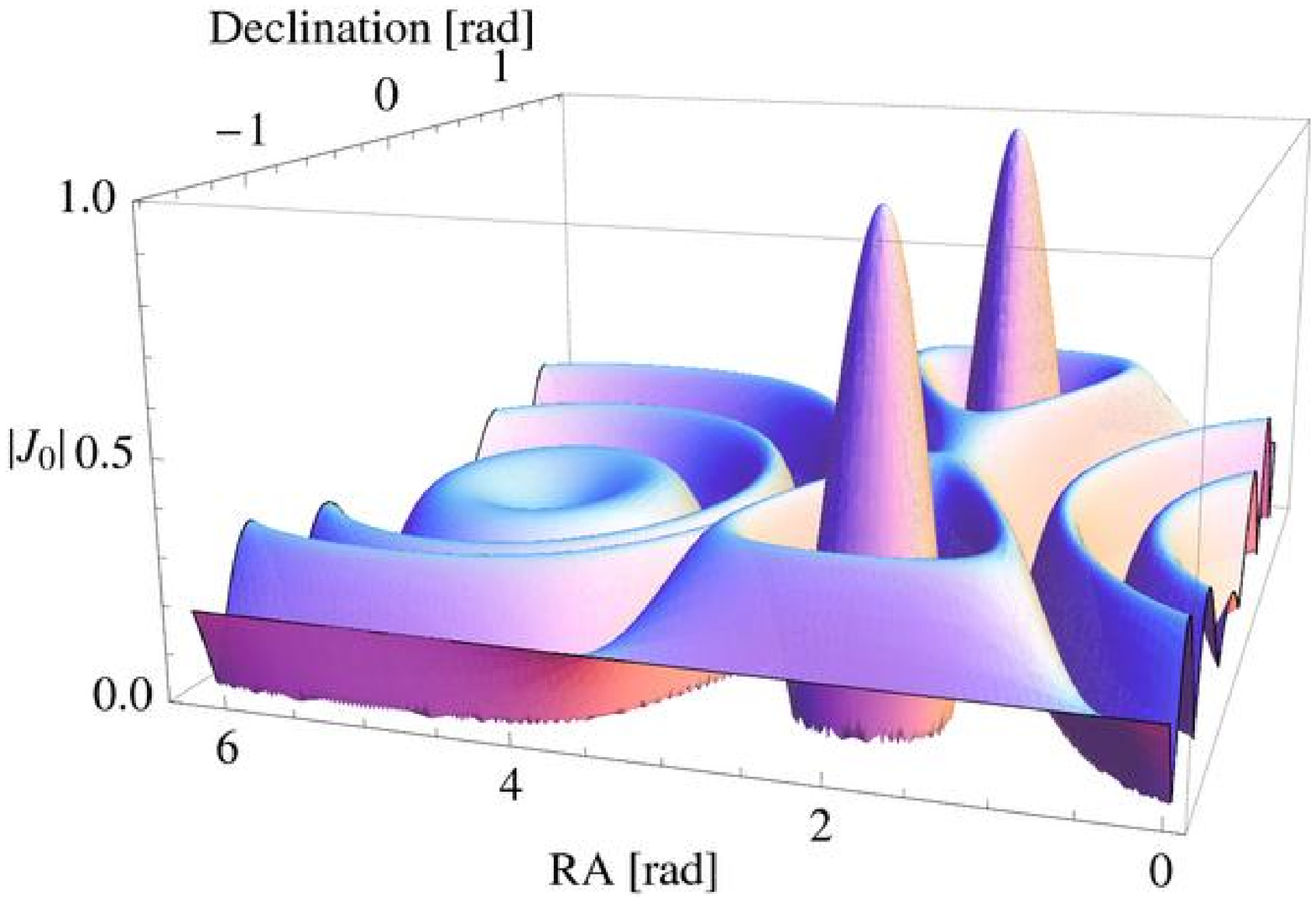}}\qquad
	\subfigure[The hardware-injected signal.\label{f:Xspin-hwinj}]{
	\includegraphics[width=0.6\columnwidth]
	{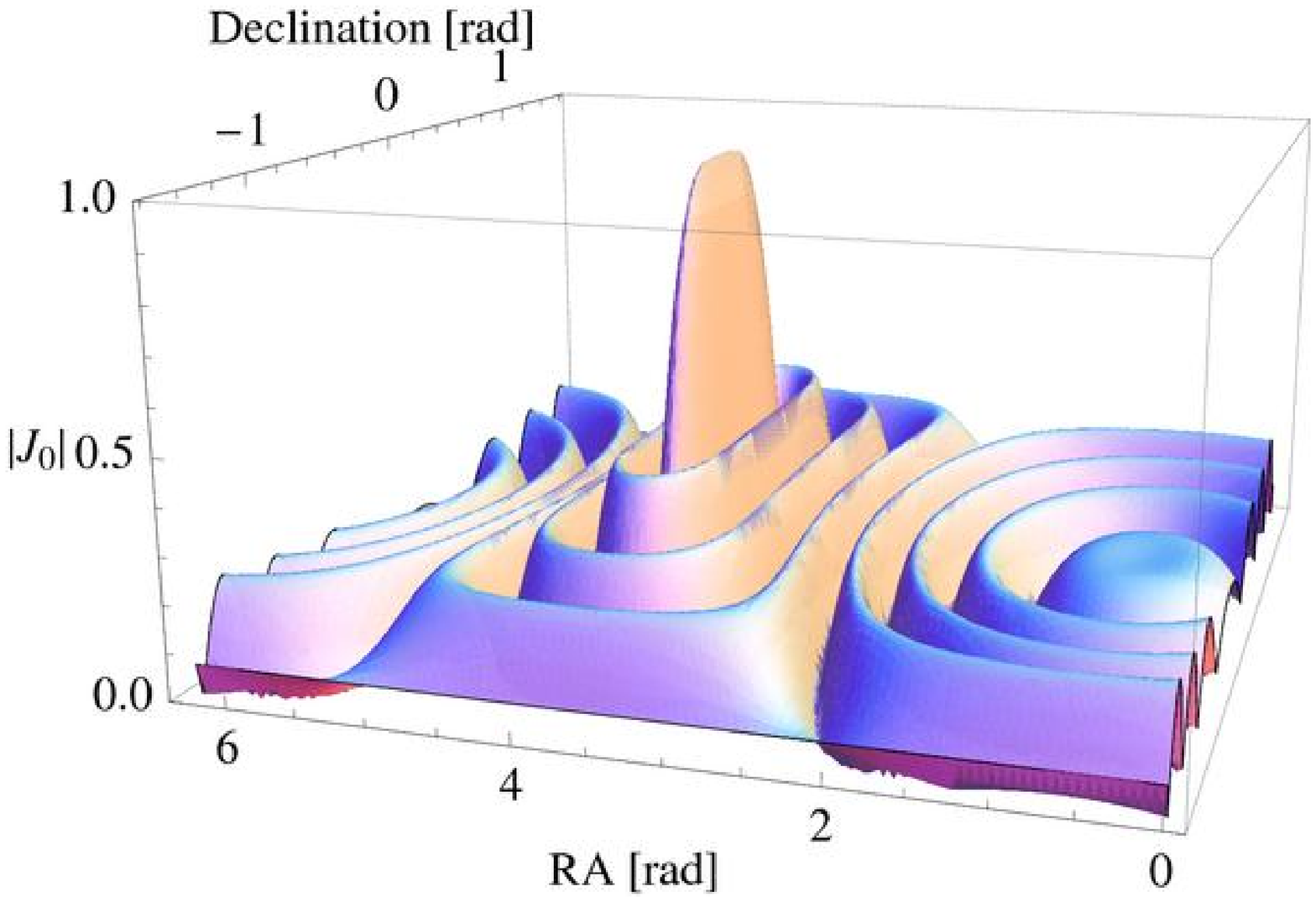}}\qquad
	\subfigure[Instrumental noise line.\label{f:Xspin-line}]{
	\includegraphics[width=0.6\columnwidth]
	{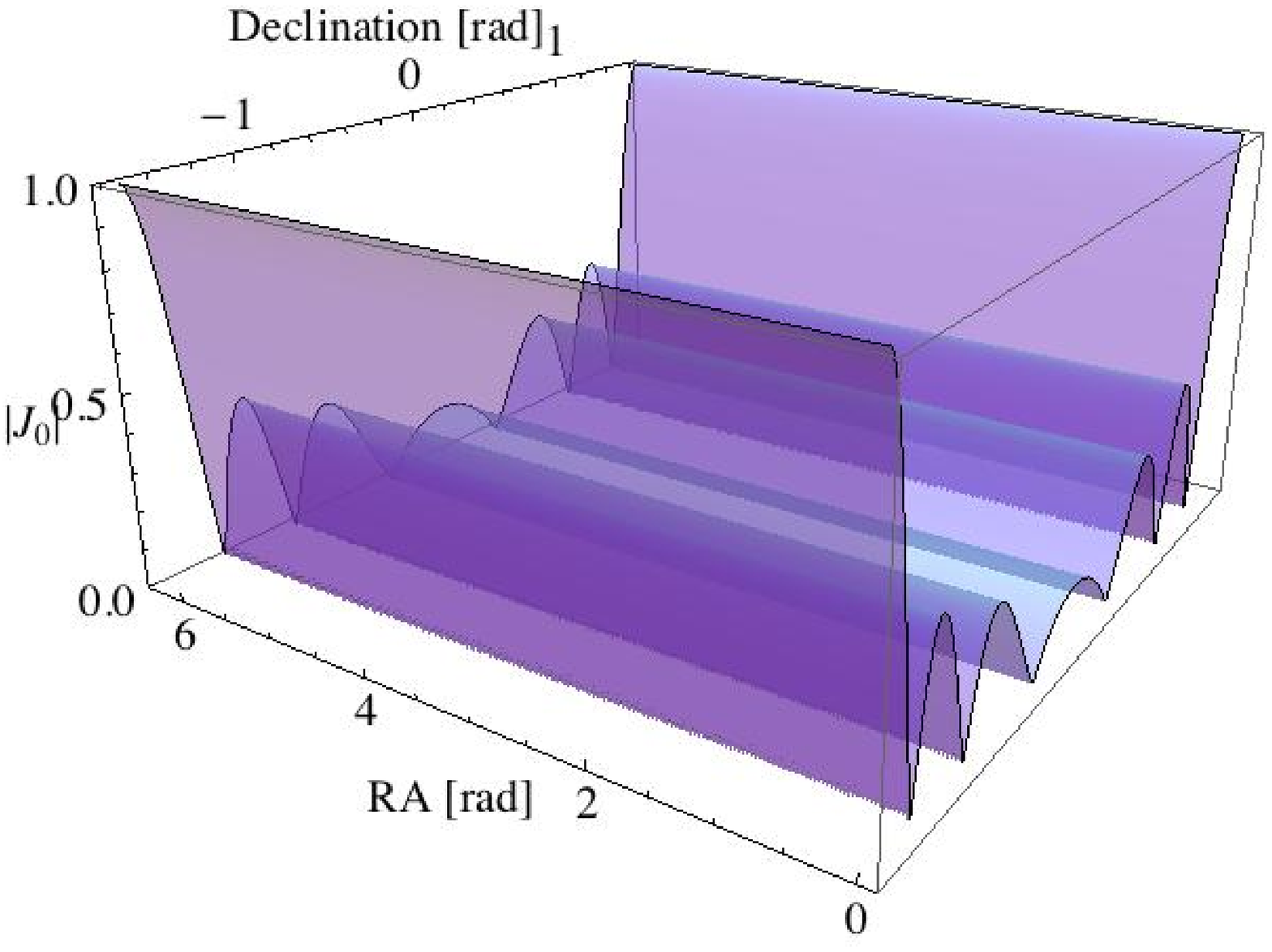}}
	\subfigure{
	\includegraphics[width=0.6\columnwidth]
	{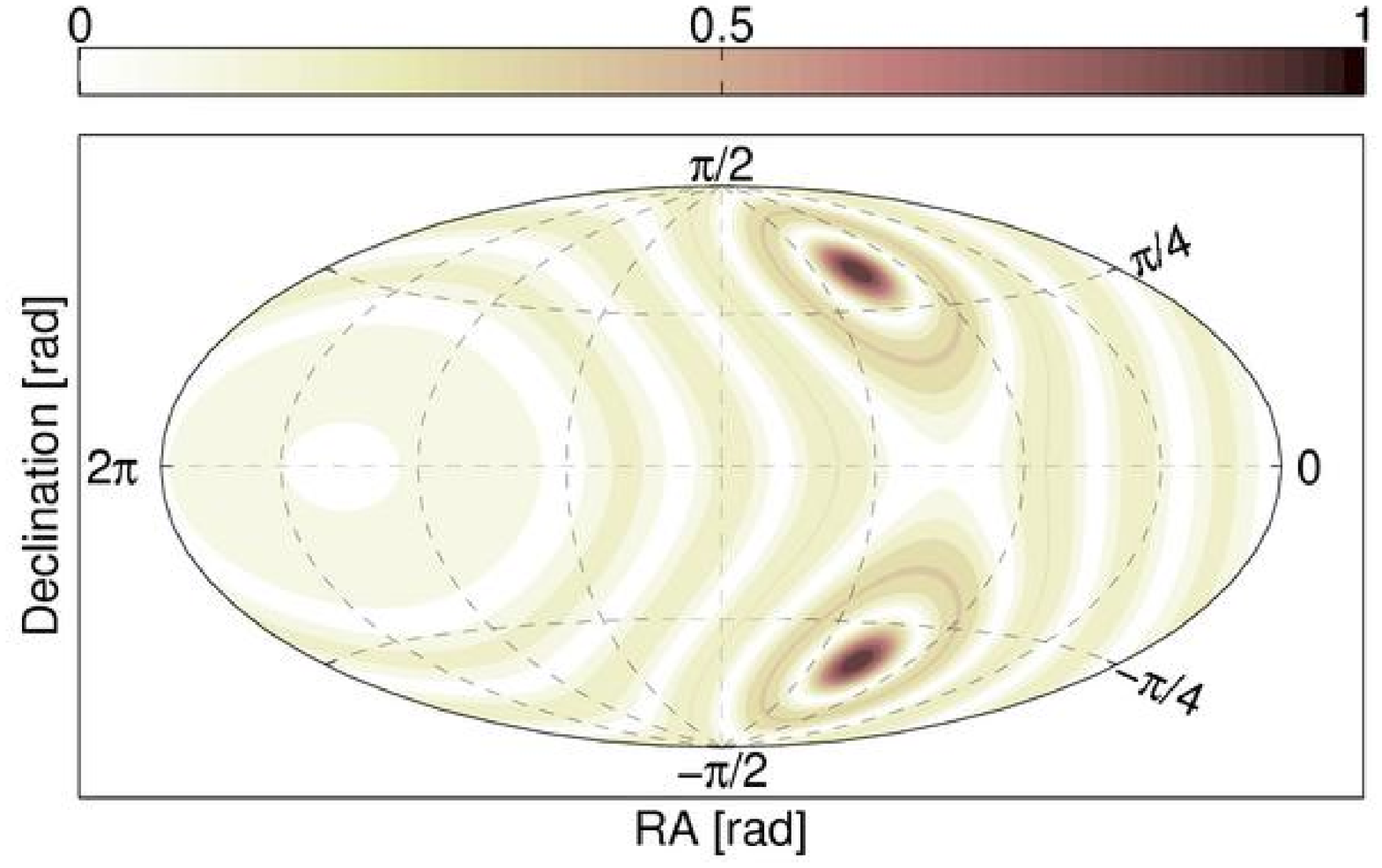}}\qquad
	\subfigure{
	\includegraphics[width=0.6\columnwidth]
	{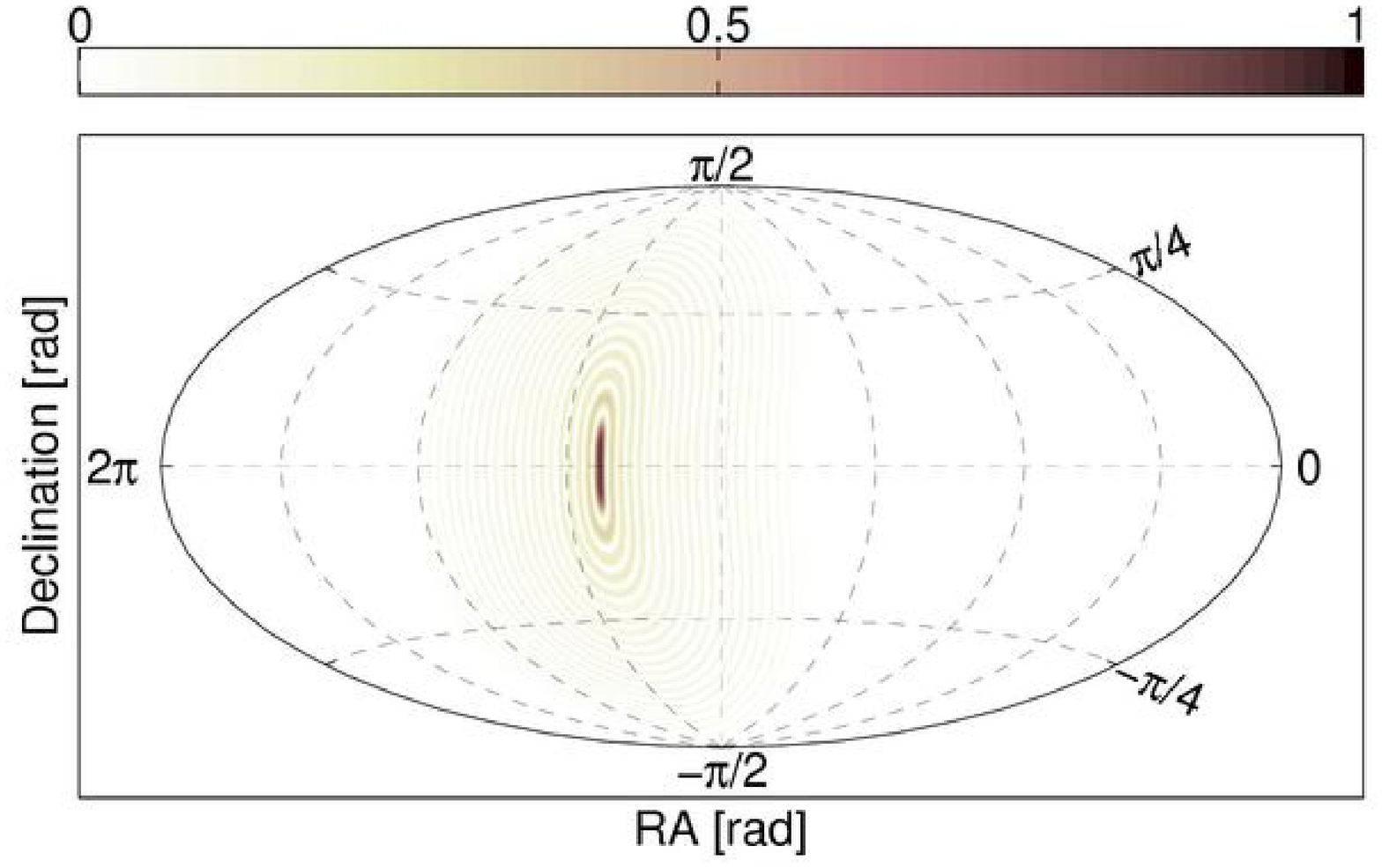}}\qquad
	\subfigure{
	\includegraphics[width=0.6\columnwidth]
	{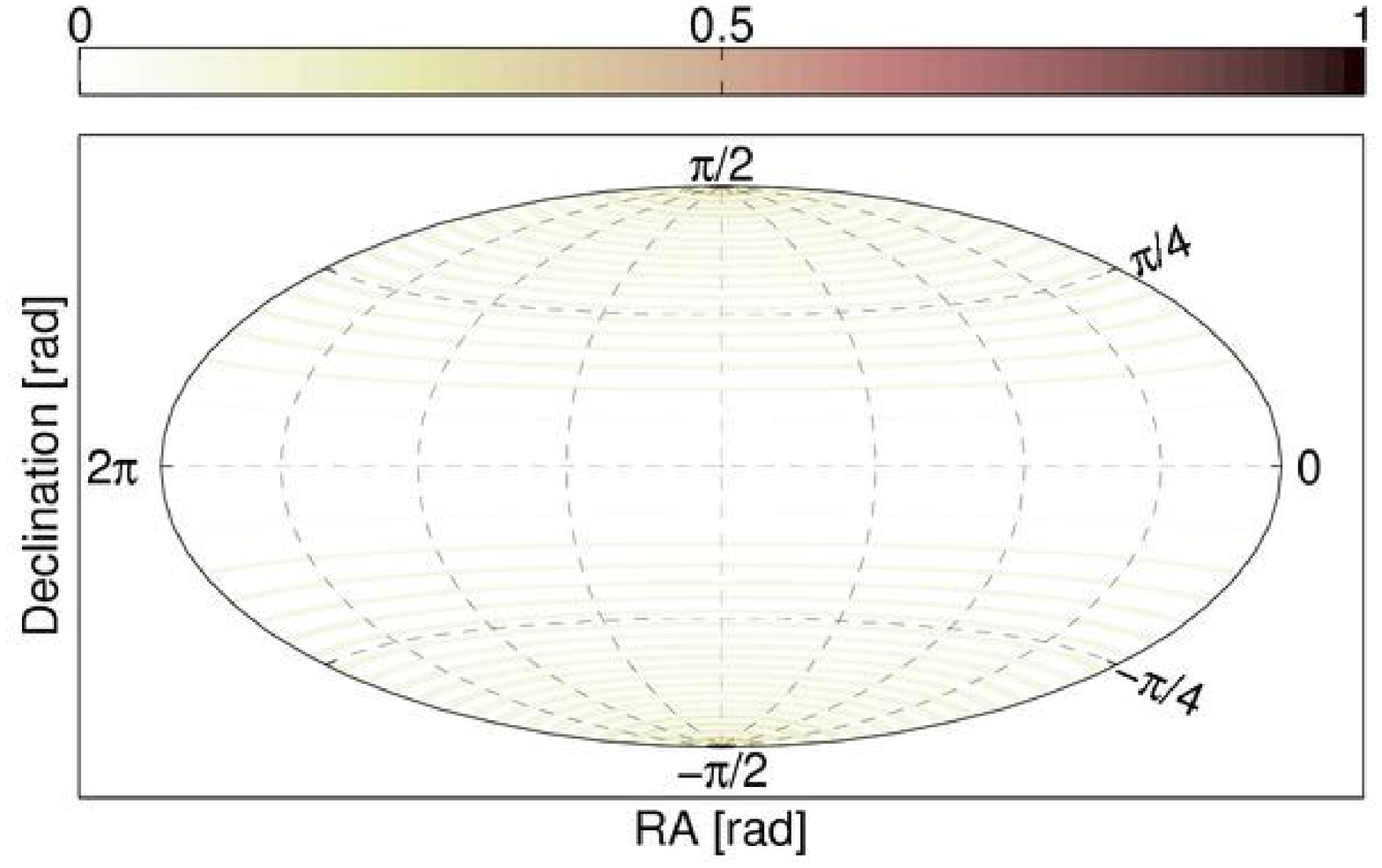}}
	\caption{(Color online) Shown is $|X_{\rm spin}| \approx |J_0( \Delta u_{\rm spin})|$ 
	over the entire sky using equatorial coordinates of
	right ascension~$\alpha$ and declination~$\delta$ for the three different examples
	studied previously in this work.
	The first column corresponds to the software-simulated signal [cf. \Fref{f:SoftwInj}],
	the second column to the hardware-injected signal [cf. \Fref{f:hwinj1}], and
	the third column to the instrumental-noise line [cf. \Fref{f:NoiseLine1}]. In the top
	row, three-dimensional plots of $|J_0( \Delta u_{\rm spin})|$ as functions of $\alpha$
	and $\delta$ are shown, and in the bottom row two-dimensional Hammer-Aitoff
	projections of the sky are given illustrating the contours of $|J_0( \Delta u_{\rm spin})|$.
	\label{f:J0}}
\end{figure*}

\Fref{f:Xspin} shows $|X_{\rm spin}|$ for the two LIGO detectors as a
function of $T$ and $\Delta u_{\rm spin}$.
It is obvious that for the observation time being an integer multiple~$\ell$ of the 
Earth's spin period, such that $T=2\pi\ell/\Omega_{\rm spin}$, \Eref{e:mod-Xspin} then
simplifies to $|X_{\rm spin}| = |J_0( \Delta u_{\rm spin})|$,
because only the term corresponding to $n=0$ does not vanish. 
By inspection, we find that this is also approximately the case
for all observation times of one day or longer, as can be seen
from \Fref{f:Xspin}. Therefore, we approximate $|X_{\rm spin}|$ for 
$T\gtrsim2\pi/\Omega_{\rm spin}$ (which is also the regime relevant to CW searches) by
\begin{equation}
  |X_{\rm spin}| \approx |J_0( \Delta u_{\rm spin})|.
  \label{e:J0}
\end{equation}
\Fref{f:J0} illustrates $|J_0( \Delta u_{\rm spin})|$ over the entire sky 
for the three different cases studied previously in this work:
the software-injected signal, the hardware-injected signal and the
instrumental-noise line.
Comparing \Fref{f:Xspin-swinj} with \Fref{f:SoftwInj}, \Fref{f:Xspin-hwinj} with \Fref{f:hwinj1},
and \Fref{f:Xspin-line} with \Fref{f:NoiseLine1}, one finds that the variation of
the $\F$-statistic in the regions determined by the global-correlation hypersurfaces
(locations consistent with $\Delta u_m=0$) can in fact be recovered in \Fref{f:J0}. 

\begin{figure*}
	\subfigure[The software-injected signal.\label{f:X-cf-cut-sw-a}]{
	\includegraphics[width=0.9\columnwidth]
	{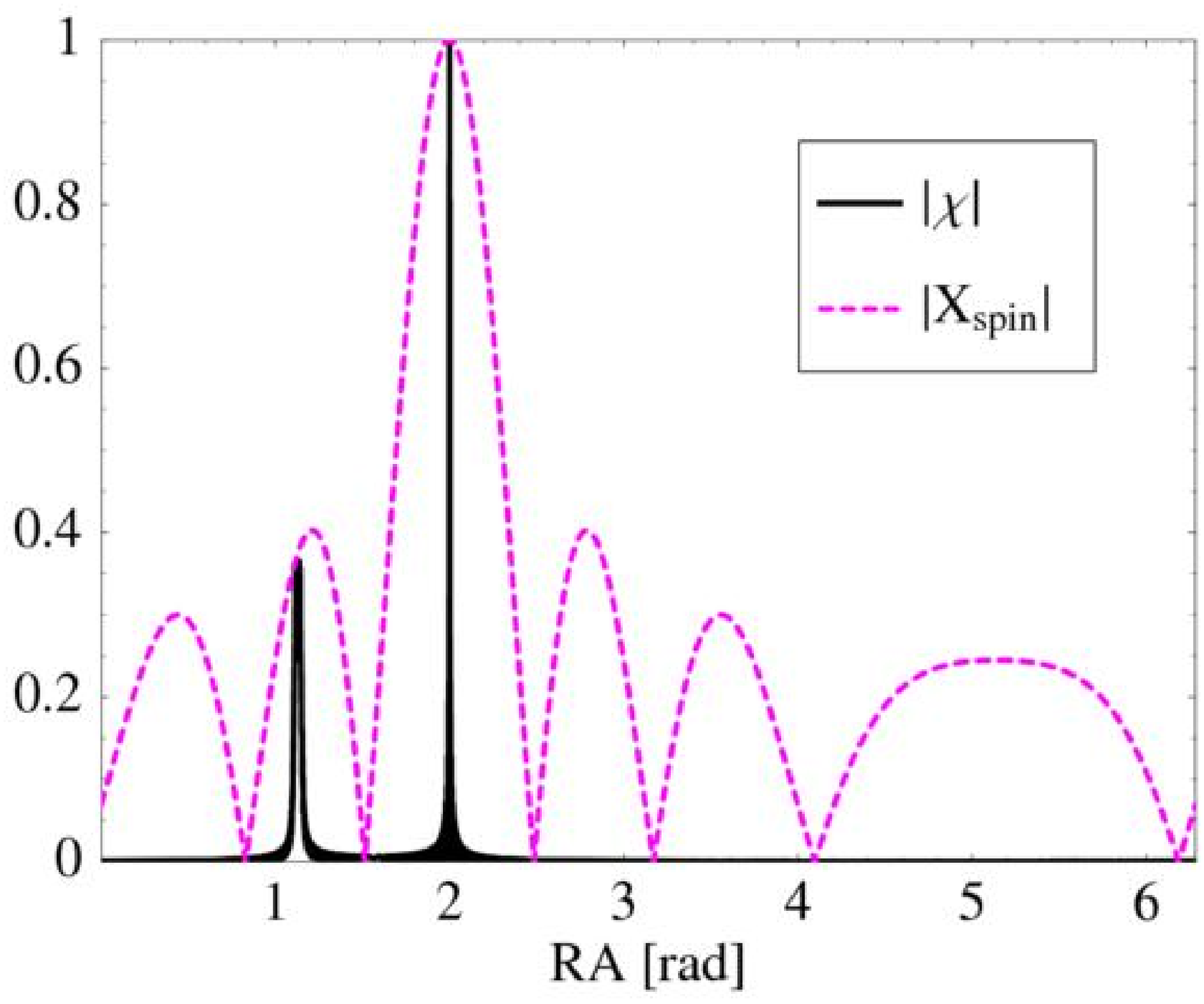}}\qquad
	\subfigure[The software-injected signal.\label{f:X-cf-cut-sw-d}]{
	\includegraphics[width=0.9\columnwidth]
	{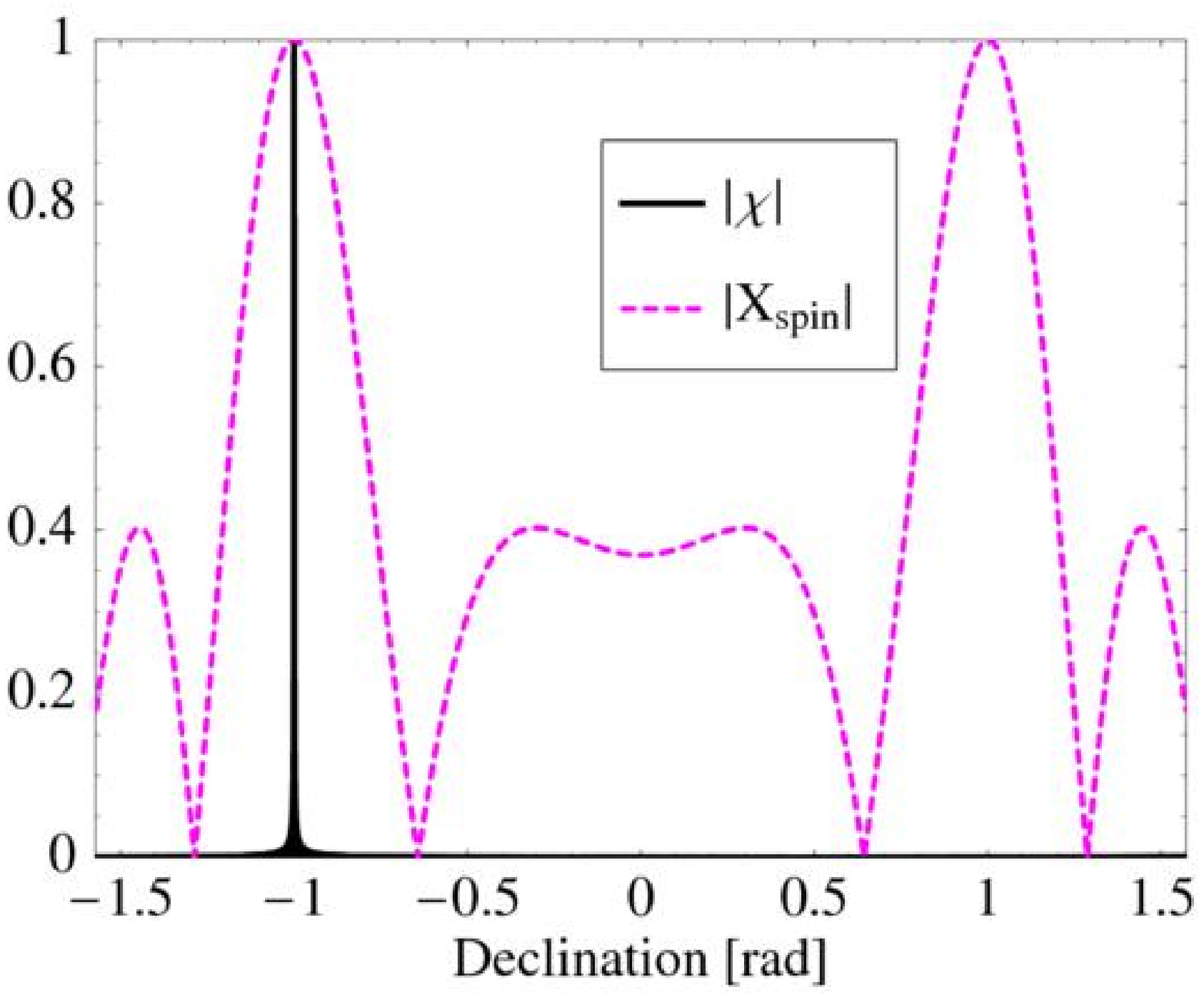}}\\
	\subfigure[The hardware-injected signal.\label{f:X-cf-cut-hw-a}]{
	\includegraphics[width=0.9\columnwidth]
	{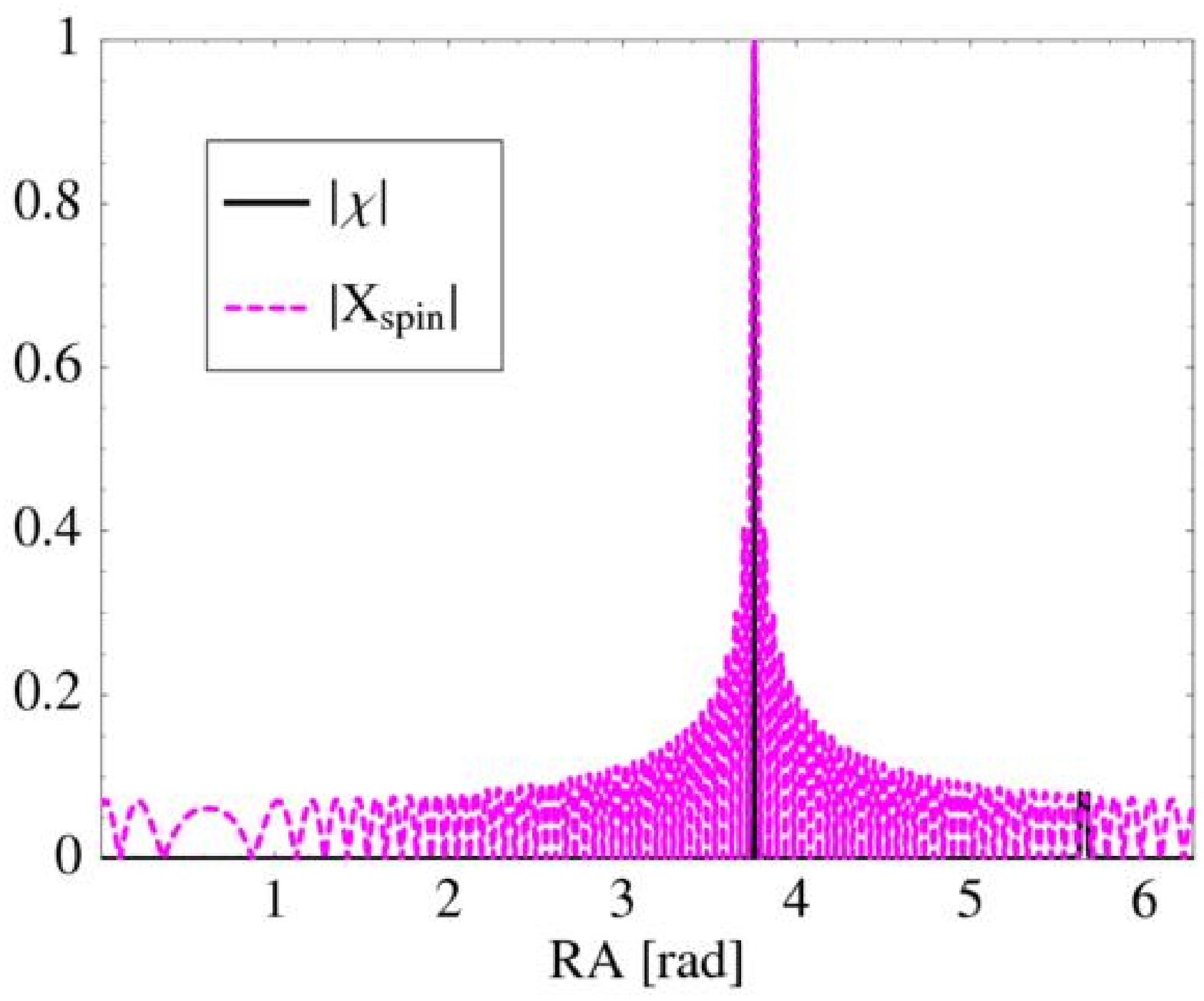}}\qquad
	\subfigure[The hardware-injected signal.\label{f:X-cf-cut-hw-d}]{
	\includegraphics[width=0.9\columnwidth]
	{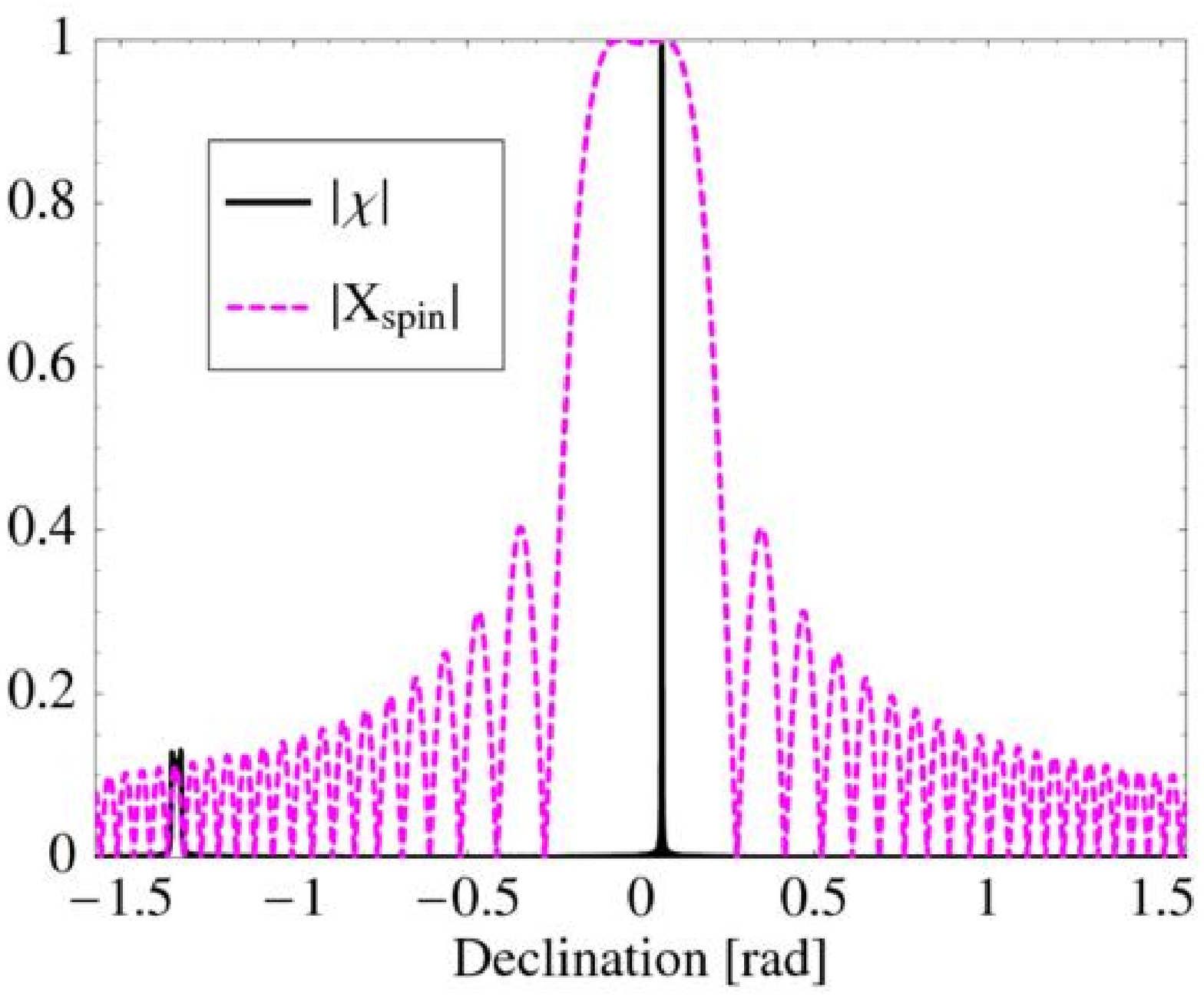}}
	\caption{(Color online) Comparison of $|\Xamp|$ (solid curves) computed using $\Phi_{\rm orb}(t)$ and 
	$|X_{\rm spin}|$ (dashed curves) calculated using $\phi_{\rm spin}(t)$ as functions of sky position. 
	In~\subref{f:X-cf-cut-sw-a} and~\subref{f:X-cf-cut-sw-d} the signal's phase parameters 
	correspond to the software injection of \Sref{ssec:SoftwareInjections}, and 
	\subref{f:X-cf-cut-hw-a} and~\subref{f:X-cf-cut-hw-d} refer to the hardware-injected 
	signal introduced in \Sref{ssec:HardwareInjection}.	
	In~\subref{f:X-cf-cut-sw-a} and~\subref{f:X-cf-cut-hw-a}, $|\Xamp|$ and $|X_{\rm spin}|$ are
	shown as functions of right-ascension (RA) template-value~$\alpha$ for the 
	given signal-value $\alpha_\sig$, while fixing the remaining template parameters to perfectly match the signal's
	parameters,  $\delta_\sig=\delta$, $f_\sig=f$ and $\fkdot{1}_\sig=\fkdot{1}$.
	Whereas in~\subref{f:X-cf-cut-sw-d} and~\subref{f:X-cf-cut-hw-d}, $|\Xamp|$ and $|X_{\rm spin}|$ are
	shown as functions of declination template-value~$\delta$, while the remaining template parameters 
	coincide with the signal's	phase parameters.
	One can see that $|\Xamp|$  decreases much more rapidly compared to
	$|X_{\rm spin}|$. Therefore $|\Xamp|$  dominantly determines  the global maximum structure
	of the detection statistic.
	\label{f:X-cf-cut}}
\end{figure*}

To illustrate better why the Earth's orbital motion determines the regions
of largest detection-statistic and the Earth's spinning motion only modulates
the detection-statistic within these regions, we consider the examples
represented in \Fref{f:X-cf-cut}.
There, as a basis of comparison to $|X_{\rm spin}|$ of \Eref{e:J0} we use $|\Xamp|$ of \Eref{e:X-m-le2}
of the orbital phase model for $m\le2$. As shown in \Sref{ssec:third-order-est}, higher orders of~$m$ are
insignificant for the example cases considered. \Fref{f:X-cf-cut} compares
$|\Xamp|$ and $|X_{\rm spin}|$ as functions of sky position for the given signal phase parameters. 
In the two diagrams~\ref{f:X-cf-cut-sw-a} and~\subref{f:X-cf-cut-hw-a}, the sky has been sliced 
along~$\alpha$ at constant declination $\delta=\delta_\sig$, and for fixed values of
frequency~$f=f_\sig$ and for simplicity also at zero spin-down offset~$\fkdot{1}_\sig=\fkdot{1}$.
In the plots~\ref{f:X-cf-cut-sw-d} and~\subref{f:X-cf-cut-hw-d}, the sky has been sliced 
along declination~$\delta$ at constant right ascension $\alpha=\alpha_\sig$, and
for the remaining template parameters coinciding with the signal's parameters.
The essential observation is that $|\Xamp|$ drops off much more rapidly compared to
$|X_{\rm spin}|$. Therefore, $|\Xamp|$ dominantly determines the global maximum structure
of the detection statistic, whereas $|X_{\rm spin}|$ only modulates
the detection statistic within these regions where $|\Xamp|$ is maximal.

\section{Conclusions}
\label{sec:conclusion}
The family of global-correlation hypersurfaces derived in this paper
provides a approximate analytical description of the global 
large-value structure of the optimal detection statistic~$\F$ in the phase-parameter
space of continuous gravitational-wave searches.

For observation times longer than one sidereal day, but still much smaller compared
to one year, it is the orbital motion of the Earth which generates a family 
of global-correlation equations.
The solution to each of these equations is a different hypersurface in parameter space.
The detection statistic is expected to have large values at the intersection of these hypersurfaces.
In this context, the Earth's spinning motion plays a minor role, because
it only varies the detection statistic within the intersection regions determined
by the global-correlation hypersurfaces.

While embedding previously published results~\cite{prixitoh:2005} in
the present theory, this work leads to a substantially
improved understanding 
of the global correlations in the optimal detection statistic.

In a comparison study with results of the $\F$-statistic from numerically 
simulated as well as from hardware-simulated signals in the presence of noise,
the analytical predictions by the global-correlation equations have been
qualitatively well recovered. 

In addition, the global large-value structure of the detection 
statistic produced by stationary instrumental-noise 
lines mimicking astrophysical sources is also well described by the 
global-correlation equations. This permitted the construction of a 
veto method, where such false candidate events are excluded.

Moreover, reparameterization of the original phase parameters by 
the parameters~$u_m$ from the global-correlation equations
offers evident advantages in solving further problems related
to CW searches. 
For example using the $u_m$-parameters can help in placing
templates more efficiently. As these parameters ``absorb" the
global correlations leading to a linear phase model, the 
metric in these parameters will be flat,
which means independent of the parameters (cf.~\cite{jk4}).

Furthermore, one could implement a candidate-event coincidence scheme via
reparameterizing the phase parameters of the candidate events by the $u_m$-parameters.
In the results from a coherent search using a bank of templates, 
a putative signal should accumulate large detection statistic values in a very focused 
region in the space of $u_m$-parameters, because for the templates located along the 
contours of the global large-value structure, the parameters $u_m$ are approximately invariant. 
In addition this fact gives rise to a novel type of  hierarchical semi-coherent 
search technique for CW sources. In such a multistage scheme~\cite{{hough:2005},{cutler:2005}} 
one breaks up the data set into a sequence of short data segments, of which each segment 
is analyzed coherently in a first stage. This is followed by an incoherent combination of 
the coherent results from each segment.
The transformation to global-correlation parameters~$u_m$ helps classifying
coincident candidate events from the first-stage coherent step before these
are combined incoherently.
However, in future work the efficiency of such a correlation transform 
scheme~\cite{pletsch:ct} as well as the relevance of the global correlations 
in the context of the template-placing problem should be investigated.

\section{Acknowledgments}
I am 	deeply grateful to Bruce Allen for his guidance and many helpful discussions. 
I also thank Reinhard Prix for stimulating conversations and useful comments on the manuscript.
I gratefully acknowledge the support of the Max-Planck-Society and the IMPRS on 
Gravitational Wave Astronomy.
I also acknowledge the LIGO Scientific Collaboration for the usage of data.

\bibliography{GPSC}

\begin{thebibliography}{18}
\bibitem[{neg()}]{negprop}
\bibinfo{note}{The effects of the neutron star's proper motion in this context
  are discussed in~\cite{jk2,jk3}. Assuming an extreme case where the star
  moves with respect to the SSB at $10^3\,{\rm km/s}$ and its distance to the
  detector to be at $40\,{\rm pc}$, over 120 days of observation, fitting
  factors (fraction of signal-to-noise ratio when using a filter not perfectly
  matching to a signal) greater than 99\% are obtained. In addition, in this
  paper observation times of order a few days, much less than 120 days, are
  considered and therefore the proper-motion effects are neglected.}
\bibitem[{ear()}]{earthdoppler}
\bibinfo{note}{The maximum Doppler frequency shift due to the Earth motion is
  given approximately by the maximum Earth's orbital velocity $v_{\rm orb}/c
  \approx 10^{-4}$.}
\end{thebibliography}

\end{document}